\documentclass[pra,aps,onecolumn,10pt,superscriptaddress]{revtex4-2}
\usepackage{amsmath,amsfonts,amssymb,graphics,graphicx,epsfig,color,times}
\usepackage[utf8x]{inputenc}
\usepackage{color}
\usepackage{natbib}
\usepackage{bbm, dsfont} 
\usepackage{subfigure}
\usepackage{mathrsfs}
\usepackage{verbatim}
\usepackage{setspace}
\usepackage{esint}
\usepackage{lineno}
\usepackage[unicode=true]
 {hyperref}
\usepackage{lineno}
\begin{document}
\title{Photon-mediated dipole-dipole interactions as a resource for quantum science and technology in cold atoms}
\author{H. H. Jen}
\email{sappyjen@gmail.com}
\affiliation{Institute of Atomic and Molecular Sciences, Academia Sinica, Taipei 10617, Taiwan}
\affiliation{Physics Division, National Center for Theoretical Sciences, Taipei 10617, Taiwan}

\date{\today}
\renewcommand{\r}{\mathbf{r}}
\newcommand{\f}{\mathbf{f}}
\renewcommand{\k}{\mathbf{k}}
\def\p{\mathbf{p}}
\def\q{\mathbf{q}}
\def\bea{\begin{eqnarray}}
\def\eea{\end{eqnarray}}
\def\ba{\begin{array}}
\def\ea{\end{array}}
\def\bdm{\begin{displaymath}}
\def\edm{\end{displaymath}}
\def\red{\color{red}}
\pacs{}
\begin{abstract}
Photon-mediated dipole-dipole interactions arise from atom-light interactions, which are universal and prevalent in a wide range of open quantum systems. This pairwise and long-range spin-exchange interaction results from multiple light scattering among the atoms. A recent surge of interests and progresses in both experiments and theories promises this core mechanism of collective interactions as a resource to study quantum science and to envision next-generation applications in quantum technology. Here we summarize recent developments in both theories and experiments, where we introduce several central theoretical approaches and focus on cooperative light scattering from small sample of free-space atoms, an atom-waveguide coupled interface that hosts the waveguide QED, and topological quantum optical platforms. The aim of this review is to manifest the essential and distinct features of collective dynamics induced by resonant dipole-dipole interactions and to reveal unprecedented opportunities in enhancing the performance or offering new applications in light manipulations, quantum metrology, quantum computations, and light harvesting innovations. 
\end{abstract}
\maketitle
\section{Introduction}

Light-matter interacting system \cite{Scully1997, Tannoudji1998, Loudon2000, Breuer2002, Walls2008, Tannoudji2011} presents one of the versatile platforms to study quantum science and offer unprecedented opportunities in quantum technology. The essential components of this system, the matter, are quantum emitters in general, which can be composed of atoms, ions, nitrogen-vacancy centers, quantum dots, superconducting qubits, and even molecules. Two essential advantages of these quantum emitters lie at the heart of their indistinguishability and their tailorable properties of intrinsic energy levels from different atomic species. The former facilitates the degenerate quantum regime as the foundation for quantum many-body physics \cite{Bloch2008, Giorgini2008}, while the latter allows the level transitions for laser cooling operation \cite{Metcalf1999, McKay2011} and trapping mechanism \cite{Eckardt2017}. It is the high controllability of these light-atom interacting quantum interfaces \cite{Hammerer2010, Lodahl2015} that makes them competitive in advancing the research fields like quantum computations \cite{Cirac1995, QIF} or quantum simulations \cite{Buluta2009, Georgescu2014, Altman2021}.

When light interacts with atoms, the universal photon-mediated dipole-dipole interactions (PMDDIs) or resonant DDIs \cite{Stephen1964, Lehmberg1970} arise due to the common electromagnetic fields experienced among the atoms. This interaction results from multiple light scattering between the atoms, leading to a pairwise and long-range interacting form. The form of this light-induced atom-atom interaction at small distance is equivalent to the mutual potential energy of two permanent dipoles \cite{Jackson1998, Zangwill2012}, whereas at long distance it becomes Coulomb-like and diminishes inversely to the mutual distance. These multiple length scales have significant effect in a macroscopic atomic ensemble and are accountable for many collective phenomena, involving cooperative radiations of superradiance \cite{Dicke1954, Gross1982} or subradiance \cite{Devoe1996}, collective Lamb shift \cite{Friedberg1973, Scully2009}, and non-equilibrium quantum dynamics. The intriguing nature of multiple scales in space and time of PMDDIs gives rise to the complexities of light-atom interacting systems, which results in the nonlinearity or the hierarchy of interactions that complicate the theoretical analysis, but can also lead to profound quantum correlations useful as a quantum resource. This strong correlation can be manifested in and built from the controlled quantum systems, which sets the foundation for studies of quantum science and for many applications enabled by quantum technology. 

In the past two decades, a growing interest and rapid developments in experiments as well as theories have spurred new directions to explore the potential of PMDDIs in quantum science and quantum technology. One of the progresses lies at well-controlled small sample preparation of atoms, which offers unprecedented opportunities to study the collective light scattering from PMDDIs \cite{Pellegrino2014, Jennewein2016, Roof2016, Jennewein2018} and allows theoretical calculations for direct comparisons. Another significant advance involves a capability of measuring long-time behaviors above noise levels \cite{Guerin2016}, which provides a deeper probe into the system dynamics and opens up new directions of nonequilibrium quantum dynamics in cold atoms. 

Theoretical developments also advance and are mainly benefited from computational power over time. Most of the central theoretical approaches have been developed before $1990$'s \cite{Scully1997, Tannoudji1998}, and they perform well especially under perturbations or away from strong coupling regimes. What restricts their prediction power, aside from the limit of computation hardware, is mainly the nature of quantum mechanics. For exact calculations on quantum particles from a bottom-up approach, an exponentially increased Hilbert space unavoidably arises and stagnates further theoretical analysis. An order of the number of state bases for a hundred two-level quantum emitters indicates $\mathcal{O}(2^{100})\approx 10^{30}$. This immense number along with its order can only be comparable to the number of sand grains on earth or the number of galaxies in our observable universe. Although this challenge remains, several attempts in theory have showcased the intriguing role of PMDDIs in light scattering beyond the widely known Beer-Lambert law \cite{Chomaz2012}, the concept of renormalization group in understanding the intricate multiple scattering of light in a dense atomic medium \cite{Andreoli2021}, an iterative method of repeated exact diagonalization of a subset of atoms in calculating scattering rates \cite{Robicheaux2020}, and machine learning approach in identifying the collective radiation behaviors \cite{Lin2022}. 

PMDDIs can be regarded as a resource to investigate fundamental quantum science in quantum optical platforms in general. To the most fundamental level, a light scattering from atoms provides one of the direct interrogations of the atomic medium, where the intensity, frequency shift, linewidth, far-field scattering directions, and correlations of light are highly related to the density, optical depth, and geometry of atoms as a whole \cite{Rehler1971}. What is more intriguing as well as intricate concerns the role of quantum correlations in light scattering processes within dense and strongly interacting atoms. It is commonly believed that up to the second-order correlations might be sufficient, which would be genuine in general when a weak coupling regime is assumed. Therefore, an open question still remains like what if light excitation was within a moderate strength or a macroscopic atomic ensemble was involved, without resorting to the exact calculations requiring system's full Hilbert space. Many other insightful directions in quantum science involve PMDDIs-assisted laser cooling, pure single photon generations via spectral shaping, optical reflectors with PMDDIs at a magic wavelength \cite{Shahmoon2017, Rui2020}, waveguide QED with infinite-range PMDDIs \cite{Chang2012, Douglas2015, Solano2017, Chang2018, Mahmoodian2018, Mahmoodian2020, Kim2021, Fayard2021, Sheremet2023, Tudela2024}, and topological quantum optics \cite{Perczel2017, Perczel2017_2, Perczel2020}. These seemingly diverse phenomena in different fundamental aspects of physics actually dwells upon one essential mechanism of light-induced dipole-dipole interactions. It is this collective atom-atom or spin-exchange interaction that manifests many distinct light or atomic properties with often unexpected enhanced performances. 
 
In the perspective of applied sciences, PMDDIs provides routes that can surpass noninteracting or single-particle regimes and allow improvements or new applications in near-term quantum technology. With synchronized spontaneously-emitted atoms inside a cavity, the superradiance laser can be made to leave behind around the $0.1$Hz linewidth limitation set by the thermal fluctuations of cavity mirrors \cite{Meiser2009}. This offers future laser sources which can be used to enhance the stability of atomic clocks \cite{Bohnet2012}. With selective radiant states by manipulating PMDDIs, an atomic array can enable an exponentially improved photon storage fidelities in an enhanced atom-light interface \cite{Garcia2017}. Recently, a proposal of using subwavelength quantum metasurfaces promises entanglement generations and makes possible other intriguing three-dimensional graph states useful for quantum information processing \cite{Bekenstein2020}. Another advancement tackles the small effects of cooperative Lamb shifts in a cubic optical lattice of atoms and enables control of these spatially-dependent shifts, leading to an improved and precise atomic clocks with better calibration \cite{Hutson2024}. Furthermore, a design of periodic and subwavelength quantum emitter rings is predicted to host superior photon transport with high efficiency as a natural light-harvesting systems \cite{Holzinger2024}. These examples present beneficial impacts to next-generation techniques and help shape the future human life, where all aspects of quantum technologies can be foreseen and advanced via PMDDIs, not limited to quantum metrology, quantum computations, and innovations in energy storage.  

\begin{figure}[t]
	\centering
	\includegraphics[width=0.98\textwidth]{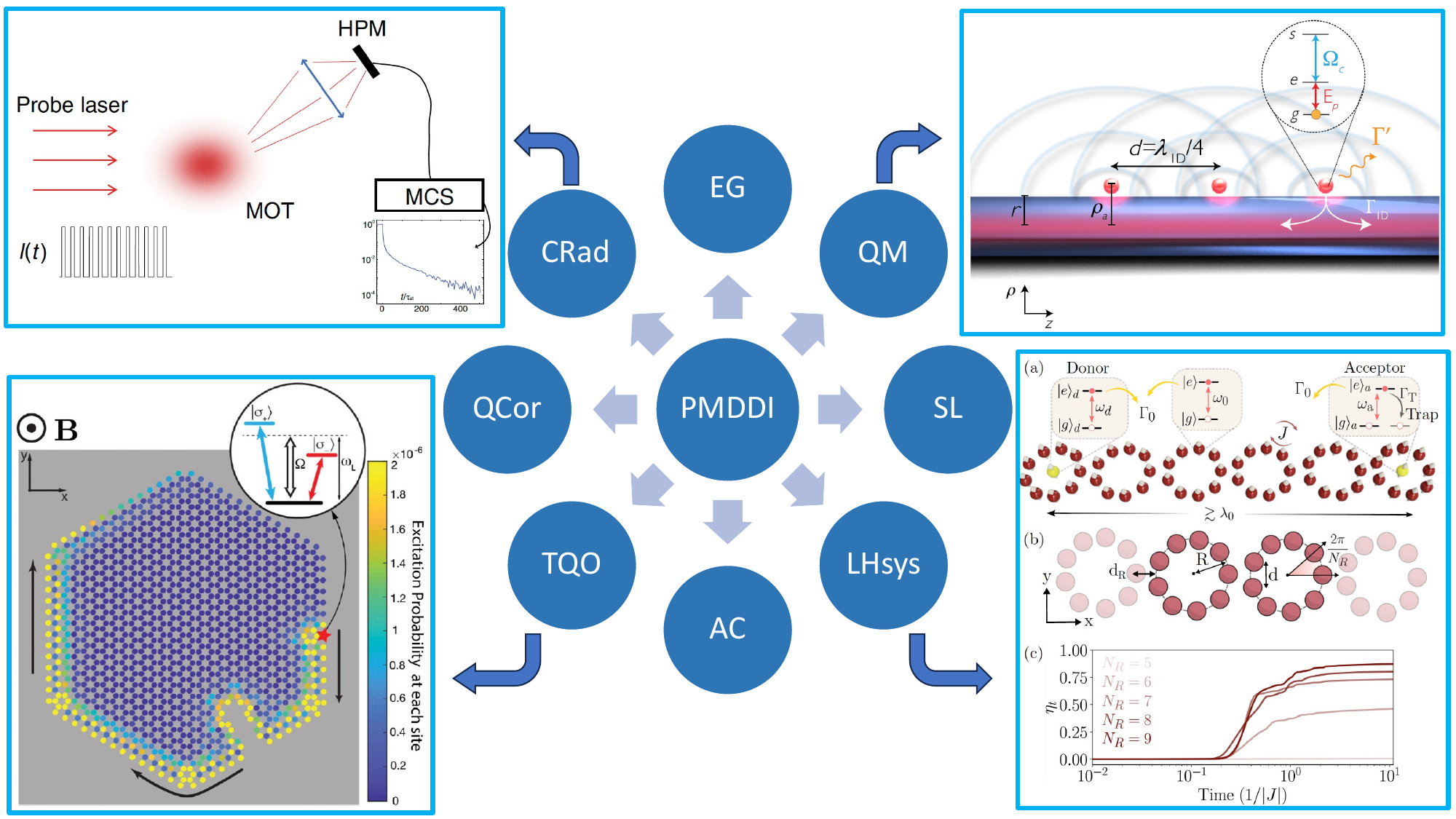}
	\caption{PMDDI as resource for quantum science and technology in cold atoms. Several selective topics include collective radiation (CRad), quantum correlation (QCor), and topological quantum optics (TQO) for studies of quantum science, and entanglement generation (EG), quantum memory (QM), superradiant laser (SL), light-harvesting system (LHsys), and atomic clock (AC) for applications in quantum technology. CRad: An experimental setup of fluorescence measurement by a hybrid photomultiplier (HPM), which is recorded on a multichannel scaler (MCS). A subradiant decay behavior is observed, a feature of collective PMDDIs. Adapted from Ref. \cite{Guerin2016}. TQO: Snapshot of the time evolution of a honeycomb lattice of atoms with an overall hexagonal shape and a large defect on one edge when an atom on the edge (red star) is driven by a laser (inset). The chosen boundaries with bearded edges support long-lived edge modes, where most of the emitted excitation is coupled into the forward direction and scattering into bulk and backward edge modes is strongly suppressed. This unidirectional energy transports around corners and routes around the large lattice defect. Adapted from Ref. \cite{Perczel2017}. QM: Electromagnetically induced transparency scheme \cite{Lukin2003, Fleischhauer2005}, where a probe field $E_p$ is coupled to the guided mode in the waveguide under a classical control field $\Omega_c$. The distance between the atoms is a quarter of the guided-mode wavelength, which allows an enhanced photon storage efficiency. Adapted from Ref. \cite{Garcia2017}. LHsys: Lattices of nanoscopic quantum emitter rings composed of (a) two-level quantum emitters with resonance frequency $\omega_0$ and a subwavelength separation $d<\lambda_0$, which are coupled via long-range dipole–dipole interactions with nearest-neighbor coupling strength $J$. Emitters acting as donor and acceptor are shown in yellow and the acceptor features an additional trapping state to which excitations irreversibly decay with rate $\Gamma_T$. (b) More detailed sketch illustrating the inter-ring separation $d_R$ and the ring radius $R$.(c) Excitation transport efficiency for a chain of $10$ rings and various number of atoms per ring $N_R$. Adapted from Ref. \cite{Holzinger2024}}\label{fig1}
\end{figure}

As shown in Fig. \ref{fig1}, this review summarizes the essential progresses in quantum science and quantum technology assisted by PMDDIs in cold atoms. We are aware that there are significant amount of relevant publications not only in the cold atom community, but also in closely related fields of quantum material and meta-materials under light excitations. Therefore, we select some of the representative works that are most relevant to each topics and are insightful to the prospects of PMDDIs as quantum resource. Moreover, this review does not cover non-Markovian regimes which consider memory effects in light and matter interactions \cite{Breuer2012, Rivas2014, Breuer2016}. For similar and tailorable long-range interacting systems \cite{Defenu2023}, exemplary platforms like trapped ions \cite{Leibfried2003, Monroe2021}, cold atomic gases in cavities \cite{Periwal2021}, dipolar and Rydberg systems \cite{Saffman2010}, and atoms near crystal waveguides \cite{Hung2016} can show many distinct phases of photon-mediated matter in addition to some distinct phenomena from PMDDIs discussed here. Lastly, in the perspective of topological phenomenon in light-matter interacting systems, topological quantum optics in two-dimensional (2D) atomic arrays \cite{Perczel2017} stands out as a distinct and an intriguing paradigm resulting from PMDDIs, which resembles polaritonic systems dressed by light \cite{Karzig2015}, but in contrast to the photonic \cite{Ozawa2019} or phononic analogues \cite{Ma2019} of topological phenomena, and the topological bands in ultracold degenerate gases \cite{Cooper2019}.  

The review is organized as follows. In Sec. II, we review theoretical approaches to investigate cold atoms with PMDDIs, which involve the effective spin-exchange interaction forms, the dipole model, light transmission beyond Beer-Lambert law, and input-output formalism. Sec. III focuses on cooperative light scattering from free-space atoms, where fundamental properties of light emissions, transmissions, or reflections are unique and sensitive to PMDDIs strengths. In Sec. IV, we summarize the recent and rapid progress on waveguide QED systems which manifest novel photonic transport, distinct quench dynamics, driven-dissipative platforms, and quantum simulations of exotic quantum many-body states. In Sec. V, we introduce topological quantum optics in 2D atomic arrays where non-Hermitian physics adds an extra ingredient to the already rich phenomena in Hermitian counterparts. Finally in Sec. VI, we conclude and discuss the outlook of PMDDIs as a resource for potential improvements and advancements in future quantum technology, as well as for opportunities in understanding better the fundamental mechanism of light-matter interactions.    

\section{Theoretical model of light-matter interacting systems}\label{Theory}

In this section, we introduce several exemplary theoretical models that reveal the essential effect of PMDDIs and can effectively describe the light response properties. In Sec. \ref{PMDDI}, we briefly review the mechanism of PMDDIs from light-matter interactions in free space and reduced dimensions with confined reservoir modes. We proceed to introduce the so-called dipole model in Sec. \ref{Dipole} which effectively takes into account of only single excitations. In Sec. \ref{transmission}, we review the light scattering properties in a small sample under significant PMDDIs. Finally we review the input-output formalism that is specifically applied in an atom-waveguide platform with long-range dipole-dipole interactions. 

\subsection{Photon-mediated dipole-dipole interactions in free space and reduced dimensions}\label{PMDDI}

A proper account of spontaneous emission from a single two-level atom requires a continuum of quantized modes in the reservoir \cite{Weisskopf1930, Scully1997}. This so-called Weisskopf-Wigner theory of spontaneous emission assumes that the emitted radiation is centered around the atomic transition frequency, which leads to an exponential decay from an excited-state atom with a Lorentzian profile in its emission spectrum. When many atoms are involved, the photon-mediated or light-induced dipole-dipole interactions manifest in atom-reservoir interactions, where the atomic excitations exchange with the common photonic modes in the reservoirs. The Hamiltonian of atom-reservoir interactions can be expressed as 
\bea
H_{\rm AR} = \sum_{\mu=1}^N \hbar\omega_{eg}\hat\sigma_\mu^\dag\hat\sigma_\mu-\sum_{\mu=1}^N\sum_{q,\lambda} g_q \left(\vec\epsilon_{q,\lambda} e^{i\k_q\cdot\r_\mu-i\omega_q t}\hat a_q +\vec\epsilon_{q,\lambda}^* e^{-i\k_q\cdot\r_\mu+i\omega_q t}\hat a_q^\dag\right)\cdot \hat d\left(\hat \sigma_\mu +\hat\sigma_\mu^\dag\right),\label{H_AR}
\eea
where the dipole approximation is assumed in the light-matter interaction described by the dipole energy of $-\vec d\cdot \vec E$ with an electric field $\vec E$ and a dipole orientation $\vec d$ with its unit direction $\hat d$ and dipole moment $d=|\vec d|$. We denote $N$ two-level alkali atoms \cite{Steck_alkali} with $|g\rangle$ and $|e\rangle$ for the ground state and the excited state, respectively, with a transition frequency $\omega_{eg}$. The atomic dipole operator is $\hat\sigma_\mu^\dag\equiv|e\rangle_\mu\langle g|$ ($\hat\sigma_\mu=(\hat\sigma_\mu^\dag)^\dag$) with $\mu$ labeling the atomic position at $\r_\mu$. The quantized bosonic fields are $\hat a_q$, satisfying the bosonic commutation relations $[\hat a_q,\hat a_{q'}^\dag]=\delta_{q,q'}$, and the coupling constant is $g_q\equiv d/\hbar\sqrt{\hbar\omega_q/(2\epsilon_0V)}$, where we consider two possible polarizations of the fields $\vec\epsilon_{q,\lambda=1,2}$ of modes $q$ and a quantization volume $V$.

Considering the Heisenberg equations $d\hat Q/dt=i[H,\hat Q]$ (let $\hbar=1$) for arbitrary atomic operators $\hat Q$ and applying the Born-Markov approximation, a master equation for $Q\equiv\langle\hat Q\rangle_0$ in Lindblad forms under the vacuum bosonic fields $\langle\rangle_0$ becomes 
\bea
\dot{Q}(t)&=&\sum_{\mu\neq\nu}^N\sum_{\nu=1}^Ni\Omega_{\mu,\nu}\left[\sigma_\mu^\dag\sigma_\nu,Q\right]+\mathcal{L}(Q),\label{Q}\\
\mathcal{L}(Q)&=&\sum_{\mu,\nu=1}^N\gamma_{\mu,\nu}\left[\sigma_\mu^\dag Q\sigma_\nu-\frac{1}{2}\left(\sigma_\mu^\dag\sigma_\nu Q+Q\sigma_\mu^\dag\sigma_\nu\right)\right].
\eea 
The $\Omega_{\mu,\nu}$ and $\gamma_{\mu,\nu}$ represent the collective frequency shifts and the decay rates respectively of the pairwise couplings, $J_{\mu,\nu}\equiv(\gamma_{\mu,\nu}+i2\Omega_{\mu,\nu})/2$, which can be defined as \cite{Stephen1964, Lehmberg1970}
\bea
\gamma_{\mu,\nu}(\xi)&=&\frac{3\Gamma}{2}\bigg\{\left[1-(\hat\p\cdot\hat{r}_{\mu\nu})^2\right]\frac{\sin\xi}{\xi}
+\left[1-3(\hat\p\cdot\hat{r}_{\mu\nu})^2\right]\left(\frac{\cos\xi}{\xi^2}-\frac{\sin\xi}{\xi^3}\right)\bigg\},\label{F}\\
\Omega_{\mu,\nu}(\xi)&=&\frac{3\Gamma}{4}\bigg\{-\Big[1-(\hat\p\cdot\hat{r}_{\mu\nu})^2\Big]\frac{\cos\xi}{\xi}
+\Big[1-3(\hat\p\cdot\hat{r}_{\mu\nu})^2\Big]
\left(\frac{\sin\xi}{\xi^2}+\frac{\cos\xi}{\xi^3}\right)\bigg\}\label{G}, 
\eea
where $\hat\p$ is parallel to the excitation field polarization, $\hat r_{\mu\nu}$ indicates a unit direction to $\hat r_{\mu}-\hat r_{\nu}$, the intrinsic decay constant is $\Gamma=d^2\omega_{eg}^3/(3\pi\hbar\epsilon_0c^3)$, and a dimensionless length scale can be denoted as an interparticle distance $\xi\equiv k_L|\r_\mu-\r_\nu|$ with $k_L=\omega_{eg}/c$.

The $\gamma_{\mu\nu}$ and $\Omega_{\mu\nu}$ in the above represent the collective decay rate and the collective Lamb shift, respectively. In the limit of long distance $\xi\gg 1$, both quantities behave as $\propto 1/\xi$ when $\hat\p\perp\hat{r}_{\mu\nu}$ or $\propto 1/\xi^2$ when $\hat\p\parallel\hat{r}_{\mu\nu}$, while at a short-distance limit $\xi\ll 1$, $\gamma_{\mu\nu}\rightarrow \Gamma$ and $\Omega_{\mu\nu}\propto 1/\xi^3$ diverges. This shows a failure in the quantum optical treatment presented here when the interparticle distance is too small, which indicates an approximate lower bound at around $\xi\sim 2\pi\times 0.1$. At such small interparticle distance around $0.1\lambda$, the atomic density is around $10^{15}$cm$^{-3}$ which approaches the quantum degenerate gas limit where atomic motional degrees of freedom become essential. This neglect of wave nature of atoms in the quantum regime puts the PMDDIs described in Eqs. (\ref{F}) and (\ref{G}) valid only when $\xi\gtrsim 0.5$. 

In a reduced 2D atomic system with 2D reservoirs, PMDDIs become even more long-ranged as $\propto 1/\sqrt{\xi}$ when $\hat\p\perp\hat{r}_{\mu\nu}$ or $\propto 1/\xi^{3/2}$ when $\hat\p\parallel\hat{r}_{\mu\nu}$ \cite{Jen2019_2D, Jen2020_book}. This similar 2D electric dyadic Green’s function has been investigated with a 2D source current in a dielectric medium \cite{Yaghjian1980}, applied in studying spatial and temporal localization of light \cite{Maximo2015}, and used to tailor the properties of superradiance \cite{Longo2016}. This effective 2D reservoir engineering can be done in trapping atoms inside a planar cavity by confining the radiation modes without the third dimension or atoms in optical tweezers on a 2D photonic waveguide \cite{Tudela2015, Tecer2024}. Meanwhile, a more commonly fabricated and intriguing platform is an atom-nanophotonic waveguide system where one-dimensional (1D) reservoirs are considered. This gives the 1D PMDDIs as 
\bea
J_{\mu,\nu}=\frac{\Gamma_{\rm 1D}}{2}\left[\cos(k_L x_{\mu,\nu})+i\sin(k_L |x_{\mu,\nu}|)\right],\label{chiral1D}
\eea 
where $\Gamma_{\rm 1D}$ $\equiv$ $2|(d/d\omega)q(\omega)|_{\omega=\omega_{eg}}g_{k_L}^2L$ \cite{Tudela2013} denotes the total decay rate. The inverse of group velocity is $|(d/d\omega)q(\omega)|$ with a resonant wave vector $q(\omega)$, the coupling strength is $g_{k_L}$, and the quantization length is $L$. Equation (\ref{chiral1D}) sets the foundation for many theoretical proposals and applications in quantum state engineering, quantum simulation, and quantum communication \cite{Lodahl2017, Sheremet2023}.  

\subsection{Dipole model and beyond}\label{Dipole}

As shown in Eq. (\ref{Q}), it quickly becomes out of reach in classical computation when the complete Hilbert space of atoms is exponentially growing as $N$ increases ($2^N$ for two-level quantum emitters). When only single excitations are considered, a reduced dimension of $N$ dipole operators is sufficient for system dynamical evolution. This so-called dipole model \cite{Arauju2017} provides an immediate and preliminary machinery to study various light scattering properties under weak excitations, which illustrates a suppressed scattering rate \cite{Pellegrino2014}, cooperative light response \cite{Jennewein2016, Roof2016, Jenkins2012, Jenkins2016, Sutherland2016_1, Sutherland2016_2, Zhu2016, Ferioli2021}, transversal subradiance \cite{Bromley2016, Jen2018_helical}, and enhanced optical cross section \cite{Bettles2016}. Recently, it is applied in modeling the collectively induced transparency of inhomogeneously broadened solid-state emitters coupled to a nanophotonic cavity \cite{Lei2023}. This shows the flexibility and convenience of the dipole model to effectively capture some essential features of complex light-matter interacting platforms.  

This dipole model comprises $N$ eigenmodes in general in the singly-excited state sector, where many intriguing and allowed subradiant modes manifest slow and long-time dynamics \cite{Scully2015, Facchinetti2016, Plankensteiner2015, Jen2016_SR, Jen2016_SR2, Sutherland2016_2}, suitable for quantum memory applications \cite{Hsiao2018}. These rich phenomena also exist in atoms in a cavity \cite{Plankensteiner2017} and metamaterial arrays \cite{Jenkins2017}. A recent theory shows fermionic behaviors in the subradiant sectors \cite{Zhang2019, Albrecht2019}, which highlights the exotic role of PMDDI in a subwavelength atomic array, leading to nontrivial correlations within the system. 

\begin{figure}[t]
	\centering
	\includegraphics[width=0.98\textwidth]{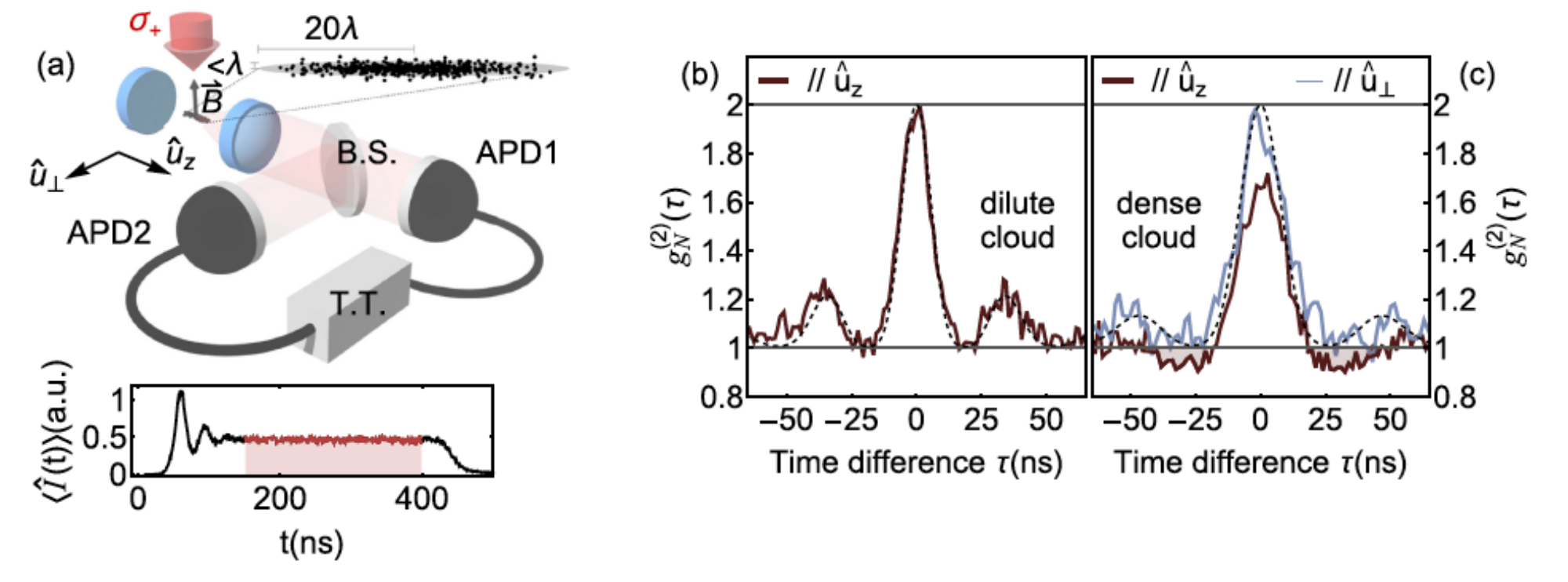}
	\caption{Experimental setup and second-order correlation measurements. (a) A cigar-shaped cloud of rubidium atoms is excited by a resonant laser beam propagating perpendicularly to its long axis. The light emitted from the cloud is collected either along its axis ($\hat u_z$) or in a perpendicular direction ($\hat u_\perp$, not shown) by two avalanche photodiodes (APD1,2). An inset plot shows an example of light intensity collected along $\hat u_z$, where red areas show the steady-state regions to calculate the correlation functions. (b) Second-order correlation functions along $\hat u_z$ for a dilute cloud, compared to the Siegert relation (dashed line). (c) Second-order correlation functions in the dense regime measured along $\hat u_z$ (red line), violating the Siegert relation. Light blue line: collection along $\hat u_\perp$. Adapted from Ref. \cite{Ferioli2024}.}\label{fig2}
\end{figure}

Going beyond the dipole model with single excitations, a sector of multiple excitations can be considered, where a restricted $C^N_M$ number of states with $M$ atomic excitations are allowed. This hugely limits the total number of states for $N$ atoms under investigation at just $M=2$ or $3$, where a less significant subradiant decay is predicted owing to a relatively larger intrinsic decay rate ($M\Gamma$) even in a subwavelength atomic array \cite{Jen2017_MP} or atomic rings \cite{Jen2018_directional}. Aside from including sectors of multiple excitations, an alternative and effective model with higher-order quantum correlations can be applied in revealing finite excitation effect \cite{Williamson2020, Robicheaux2021,  Robicheaux2023} or showing superradiant dynamics \cite{Bigorda2023, Kusmierek2023}. 

A recent experiment on steady-state second-order correlations reveals the non-Gaussian correlations in dense atomic clouds \cite{Ferioli2024}. As shown in Fig. \ref{fig2}, a clear violation of Siegert relation, relating the second-order correlations to the first-order ones based on Gaussian chaotic light, is demonstrated due to the presence of higher-order correlations. This presents a feature of a correlated medium where higher-order or even all-order atom-atom correlations induced by collective PMDDIs are needed to be accounted for, especially in a dense medium, and as well demands a theoretical framework beyond Gaussian properties \cite{Funo2024}. An exact diagonalization approach in studying light transmission property in a few of dense atoms also manifests the essence of PMDDIs, where significant deviations emerge in the mean-field approach compared to the exact model due to the ignorance of high-order correlations \cite{Wang2024}.   

\subsection{Light transmission}\label{transmission}

\begin{figure}[t]
	\centering
	\includegraphics[width=0.98\textwidth]{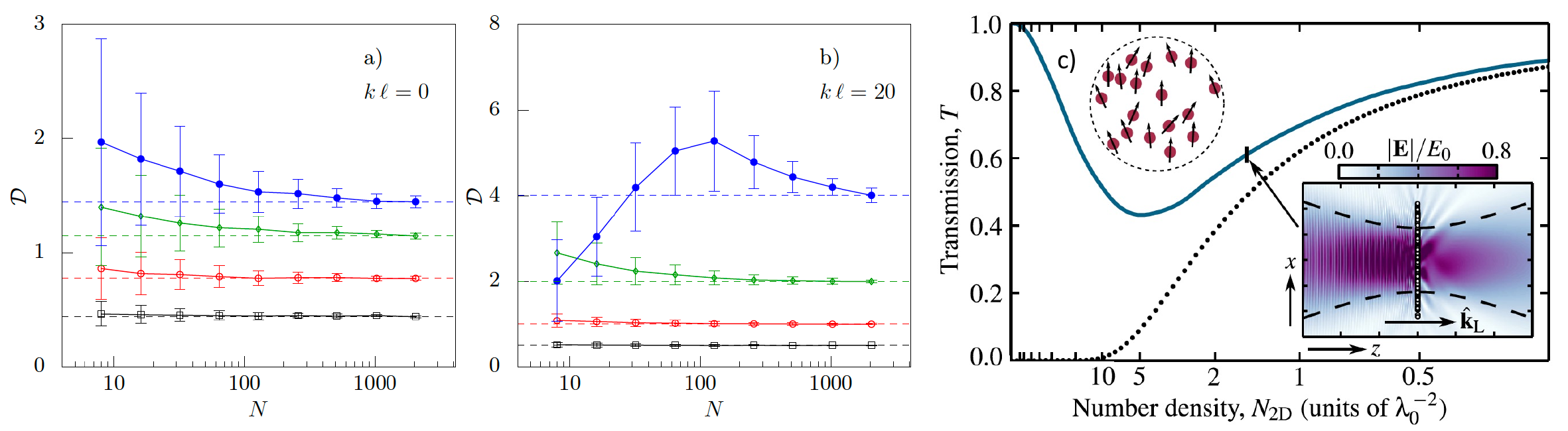}
	\caption{Variation of the optical depth $\mathcal{D}=-\ln|\mathcal{T}|^2$ and light transmission $\mathcal{T}$. In (a) and (b), $\mathcal{D}$ is obtained under resonant excitations for various number of atoms $N$ at different ensemble lengths $\it{l}$ at $\sigma_0 n^{\rm(2D)}$ = 0.5 (black), 1 (red), 2 (green) and 4 (blue) with $\sigma_0\equiv 3\lambda^2/(2\pi)$. The bars indicate the standard deviations from random distributions of atoms. The dotted lines are given from the case at $N = 2048$. Adapted from Ref. \cite{Chomaz2012}. In (c),  a resonant optical transmission of a Gaussian beam along $\hat k_L$ through a random 2D monolayer of $100$ interacting dipoles is shown. As the 2D number density increases, the interacting monolayer (blue solid line) deviates from the one with noninteracting dipoles (black dotted line). Each data point is averaged over $100$ realizations, and $x$ and $z$ vary between $\pm 6\lambda$ and $\pm 30\lambda$, respectively. The black dashed line shows the $1/e$ beam width and the white circles the atom positions. Adapted from Ref. \cite{Bettles2016}.}\label{fig3}
\end{figure}

Aside from the dipole model we introduce in the previous section, which treats the atoms as a whole being influenced by PMDDIs, a light transmission in the far field can provide further insights of atomic optical depths that can be directly measured and compared with theories \cite{Corman2017}. The optical depth can be defined as $-\ln|\mathcal{T}|^2$ with light transmission $\mathcal{T}$ measured in the far field as \cite{Chomaz2012, Bettles2016, Wang2024}  
\bea
\mathcal{T}=1+i\frac{3\Gamma}{\Omega_0R^2k^2}\sum_{\mu=1}^N \langle \sigma_\mu \rangle e^{-ikz_\mu},
\eea
where $k=2\pi/\lambda$ denotes a wave vector with a resonant wavelength $\lambda$, $\Gamma$ quantifies the intrinsic spontaneous decay rate, $\Omega_0$ denotes the input energy, and $R$ represents a radius of a cylinder geometry relating to the atomic column density $n^{\rm (2D)}=N/(\pi R^2)$ with uniform distribution. This presents a global probing scheme \cite{Chomaz2012} as shown in Fig. \ref{fig3}, where significant deviations of optical depths or light transmission are predicted either when more atoms are involved at a fixed atomic column density or when the density increases at a fixed $N$. This presents the essential role of PMDDIs in modifying the light propagation properties, very different from the Beer-Lambert law assuming a medium with independent dipoles \cite{Chomaz2012}. In particular, in Fig. \ref{fig3}(c), a transparent light emerges at an extremely dense atomic cloud due to large collective frequency shifts, in huge contrast to the noninteracting dipole model that predicts extinction of light instead \cite{Bettles2016}. 

Furthermore, when a structured 2D atomic array is considered, the cooperative resonances of the surface modes can lead to a complete reflection of light or an extinction of light transmission. This perfect optical reflector emerges at some magic lattice spacings $a/\lambda=0.2$ and $0.8$ \cite{Shahmoon2017}, which allows an efficient optical mirror by tailoring the atom density and holds promises to optical metamaterial engineering using structured ensembles of atom \cite{Rui2020}. Several recent theoretical studies in 2D atomic arrays involve light propagation in two 2D lattices \cite{Javanainen2019}, entanglement generations \cite{Ballantine2021}, alternative quantum computing platforms \cite{Shah2024}, and light reflection and transmission properties \cite{Robicheaux2024}. 

\subsection{Input-output formalism}\label{IO}

\begin{figure}[t]
	\centering
	\includegraphics[width=0.7\textwidth]{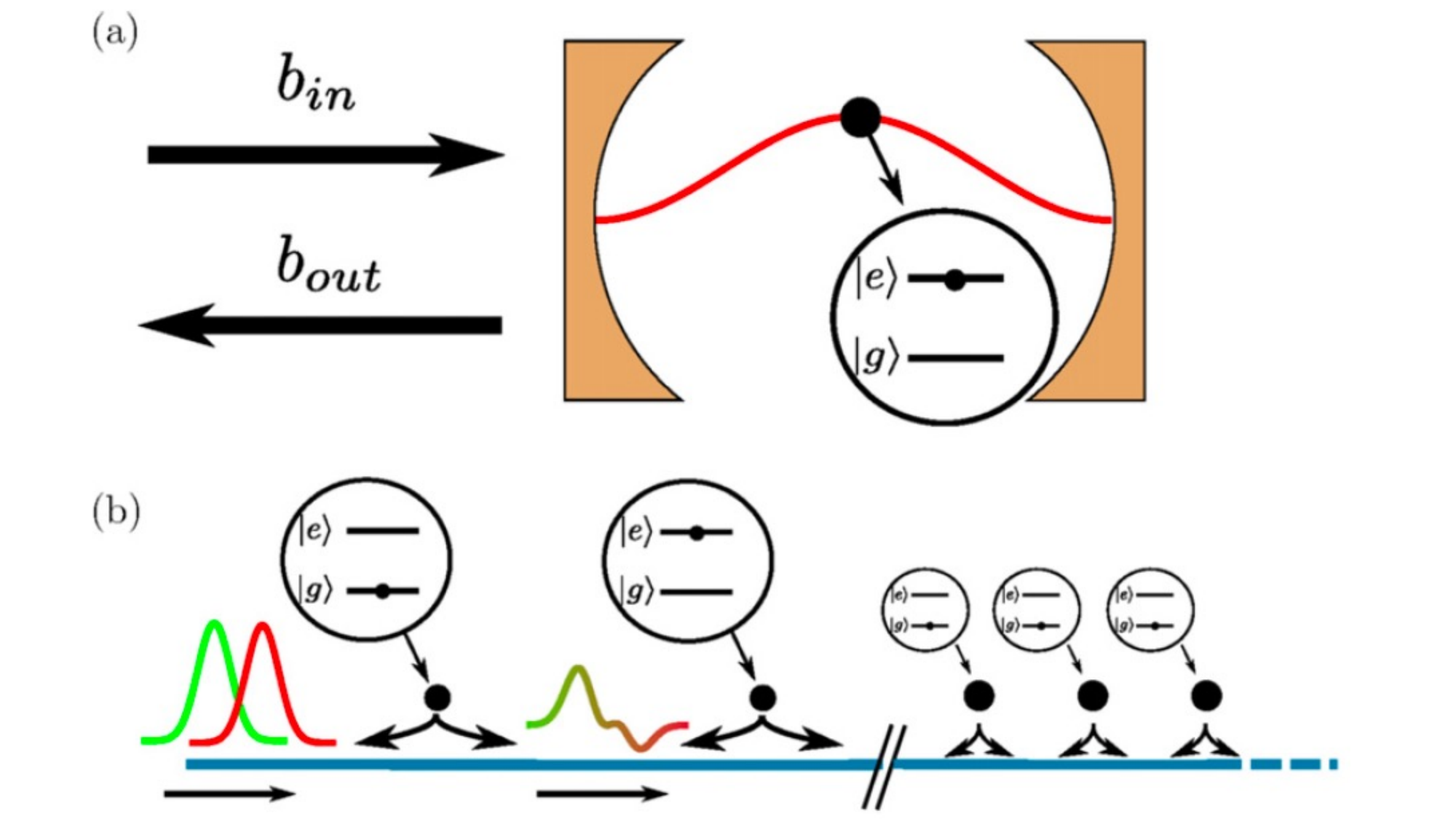}
	\caption{A cavity QED scheme and many-atom waveguide coupled platform. (a) A cavity QED system presents an input–output formalism that relates the outgoing field to the input field and the internal dynamics of the cavity system. (b) Photon scattering through the many-atom system coupled to the guided mode supported on a waveguide. The nonlinearity in photon scattering emerged from multiple scattering events among the many-atom system becomes more complicated than the cavity QED platform owing to the collective PMDDIs. Adapted from Ref. \cite{Caneva2015}.}\label{fig_in_out}
\end{figure}

In this last subsection of theoretical models on PMDDIs, we review the input-output formalism in Fig. \ref{fig_in_out} that is originally discussed in quantum optical setups with an atom inside a cavity \cite{Walls2008, Meystre2007}. In this seminal consideration, an output field can be linked to an input field in a single-sided cavity $a_{\rm out}(t)+a_{\rm in}(t)=\sqrt{\kappa}a(t)$, where $g^2=\kappa/(2\pi)$ relates a Markov quantum stochastic process $\propto \kappa$ to the reservoir couplings $g$ of the intra-cavity fields $a$, which evolves as
\bea
\dot{a}(t)=i[H_{\rm sys}, a]-\frac{\kappa}{2}a(t)+\sqrt{\kappa}a_{\rm in}(t), 
\eea
with $H_{\rm sys}$ denoting a system Hamiltonian. For a generalized input-output formalism in an atom-nanophotonic platform with multiple atoms coupled to the guided light modes, $H_{\rm sys}$ can be effectively described by $H_{\rm eff}$ under the weak field approximation where atoms are barely excited ($\sigma_{gg}\approx 1$) \cite{Caneva2015}, 
\bea
H_{\rm eff}= H_{\rm at}-\frac{i\Gamma_{\rm 1D}}{2}\sum_{\mu=1}^N\sum_{\nu=1}^N\sigma_\mu^\dag\sigma_\nu e^{ik_{\rm in}|x_\mu-x_\nu|},\label{eff}
\eea
with an intrinsic atomic Hamiltonian $H_{\rm at}$ and single-atom coupling rate to the waveguide modes $\Gamma_{\rm 1D}$. The master equation for the atomic density matrix $\rho$ can be obtained by $\dot\rho=-i[H_{\rm eff}, \rho]$, and the second term in Eq. (\ref{eff}) with spin-exchange interactions shows the exact form of PMDDI in reduced dimensions in the singly-excited atomic Hilbert space. 

The generalized input-output formalism can be cast into a form of $S$-matrix interpretation which contains complete information of photon transport. The $S$-matrix has been used to calculate the few-photon transport in a one-dimensional nanophotonic waveguide coupled to a single atom \cite{Fan2010, Shi2011, Pletyukhov2012, Xu2015}. A relation to $S$-matrix elements from the input-output formalism can be derived from the atomic dynamics determined by $H_{\rm eff}$, which further leads to many-atom cases with either chiral \cite{Jones2020} or reciprocal coupling channels, providing a useful method to study nonlinear photon transport \cite{Caneva2015}.

\section{Cooperative light scattering}

Light scattering presents one of the pristine experimental demonstrations which can in principle be directly compared with theories. But it turns out unexpectedly that many distinct and unresolvable comparisons between theories and experiments have led to demands of more sophisticated theoretical models to better describe experimental results. In particular for dense and cold atom clouds, multiple scattering effect becomes essential, and intriguing deviations from mean-field models further manifest the unavoidable roles of quantum correlations in light scattering. Here I focus on two parameter regimes with weak or finite light excitations, which highly relate to the mechanism of emerging correlations among the atoms.   

\subsection{Weak excitations}

An assumption of weak excitation is benign for theoretical studies, since a finite Hilbert space sector with singly-excited atomic states would be sufficient for theoretical analyses. In this regime, a dipole model in Sec. \ref{Dipole} is sufficient to compare with experimental observations and can predict many unique cooperative phenomena not seen in noninteracting regimes. One natural and straightforward construction of singly-excited states is termed as timed Dicke state \cite{Scully2006, Mazets2007}, which preserves the symmetry of exchanges among every pair of atoms. This owes to triplet states in two spin-$1/2$ systems and to the symmetric sectors of Dicke Hilbert space as well with a maximal total spin $J=N/2$ in $N$ spin-$1/2$ particles. On single photon absorption, the timed Dicke state can be formed and shows an enhanced decaying behavior as superradiance. Other $N-1$ non-symmetric and orthogonal states can be constructed \cite{Mazets2007}, where some of them showcase the long-time decays as an afterglow of superradiance \cite{Mazets2007}.   

\begin{figure}[t]
	\centering
	\includegraphics[width=0.7\textwidth]{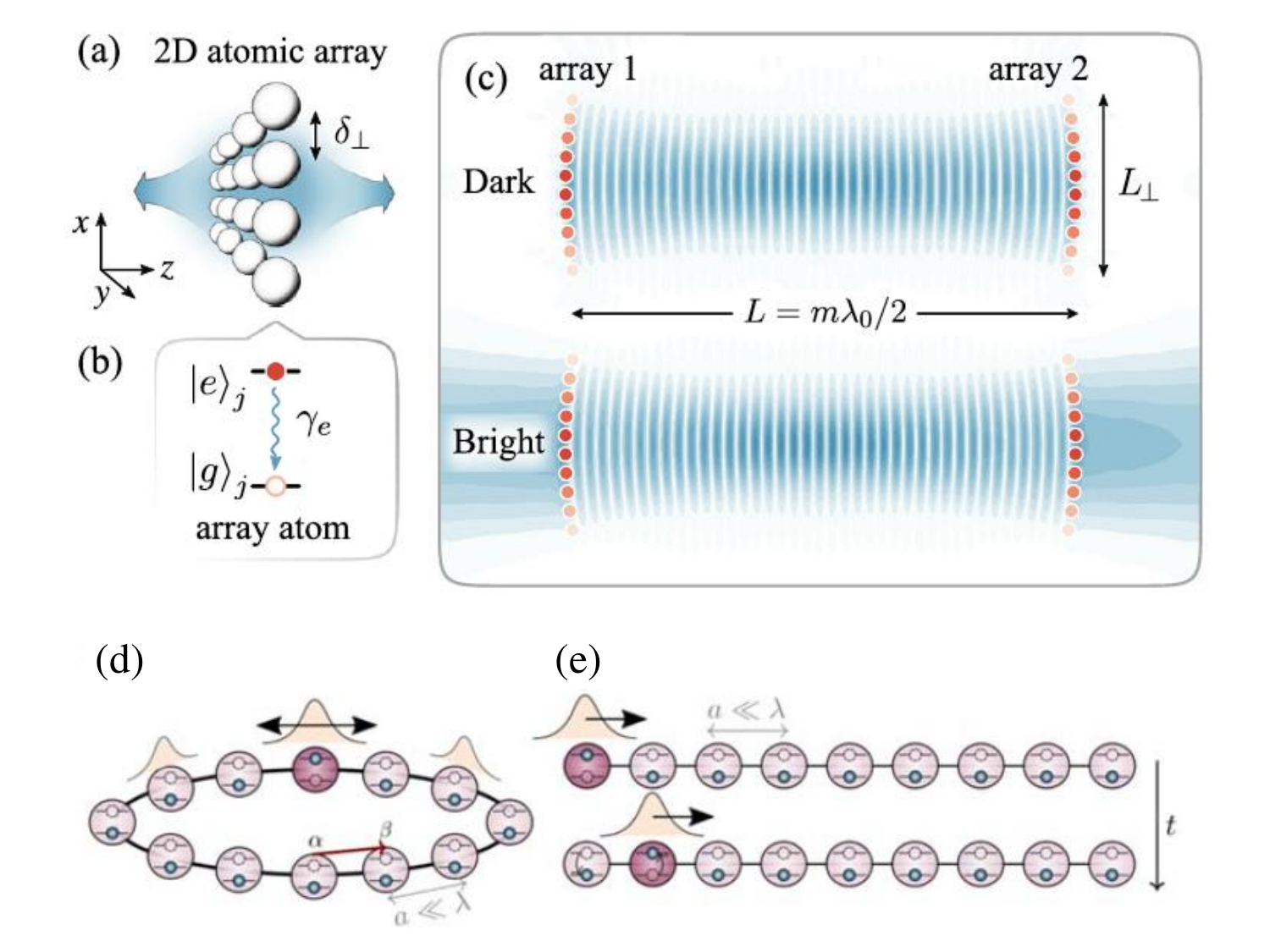}
	\caption{Subradiant state manipulations and excitation transport. (a) Schematic 2D atomic array, with light emitted perpendicular to the atomic plane in a bright or radiating state. (b) Two-level configuration. (c) Setup with two distant atomic arrays, where the electric field profile of photonic modes (blue) associated with the dark and bright states as excitations in the two arrays (red). Adapted from Ref. \cite{Guimond2019}. A 1D atomic chain with nearest neighbor separation $a$ in (d) a ring and (e) an open linear chain. The wave packet indicating the excitation is transported via PMDDIs. Adapted from Ref. \cite{Needham2019}.}\label{weak}
\end{figure}

Alongside the superradiance, the associated frequency shifts \cite{Friedberg1973, Scully2009, Jen2015, Hsu2024} have been observed in iron atoms embedded in a planar cavity \cite{Rohlsberger2010}, vapor cell \cite{Keaveney2012, Peyrot2018}, finite ionic chains \cite{Meir2014}, atomic ensembles \cite{Pellegrino2014, Jennewein2016, Jenkins2016}, and superconducting qubits \cite{Wen2019}. This manifests the strong PMDDIs induced by collective light-matter interactions, which would otherwise be vanishing as in single particles. This frequency shift is also termed as ``collective Lamb shift" to differentiate from the Lamb shift due to vacuum fluctuations. Collective light-matter interaction also leads to subradiant radiations in a dense cloud \cite{Guerin2016, Bromley2016, Zhu2016, Shahmoon2017, Hebenstreit2017, Parmee2020} and transversal light scattering orthogonal to excitation light in atomic ring structures \cite{Jen2018_SR1, Jen2018_SR2, Moreno-Cardoner2019, Moreno-Cardoner2022}. Several proposals to manipulate subradiant states \cite{Scully2015, Facchinetti_2016, Jen2016_SR, Jen2016_SR2, Sutherland2016_2, Bettles2016, Jenkins2017, Jen2017_MP, Bhatti2018, Ferioli2021_sub} involve phase imprinting \cite{Jen2016_SR}, selective radiance \cite{Garcia2017}, and collective anti-resonances from atomic arrays in a cavity \cite{Plankensteiner2017}. Subradiant states can also be prepared in separated atomic arrays \cite{Guimond2019} as shown in Figs. \ref{weak}(a-c) and fermionic atoms in optical lattices \cite{Pineiro2019}. As an application for example in Figs. \ref{weak}(d-e), a transport of subradiant excitation can be conducted in an open linear chain of atoms or an atomic ring to achieve lossless energy transport or storage \cite{Needham2019}. We expect many more applications in quantum state engineering and quantum transport with PMDDIs, and these fruitful theoretical proposals and experimental observations also provide opportunities for fundamental studies of many-body physics in light-matter interactions.  

\begin{figure}[tb]
	\centering
	\includegraphics[width=0.85\textwidth]{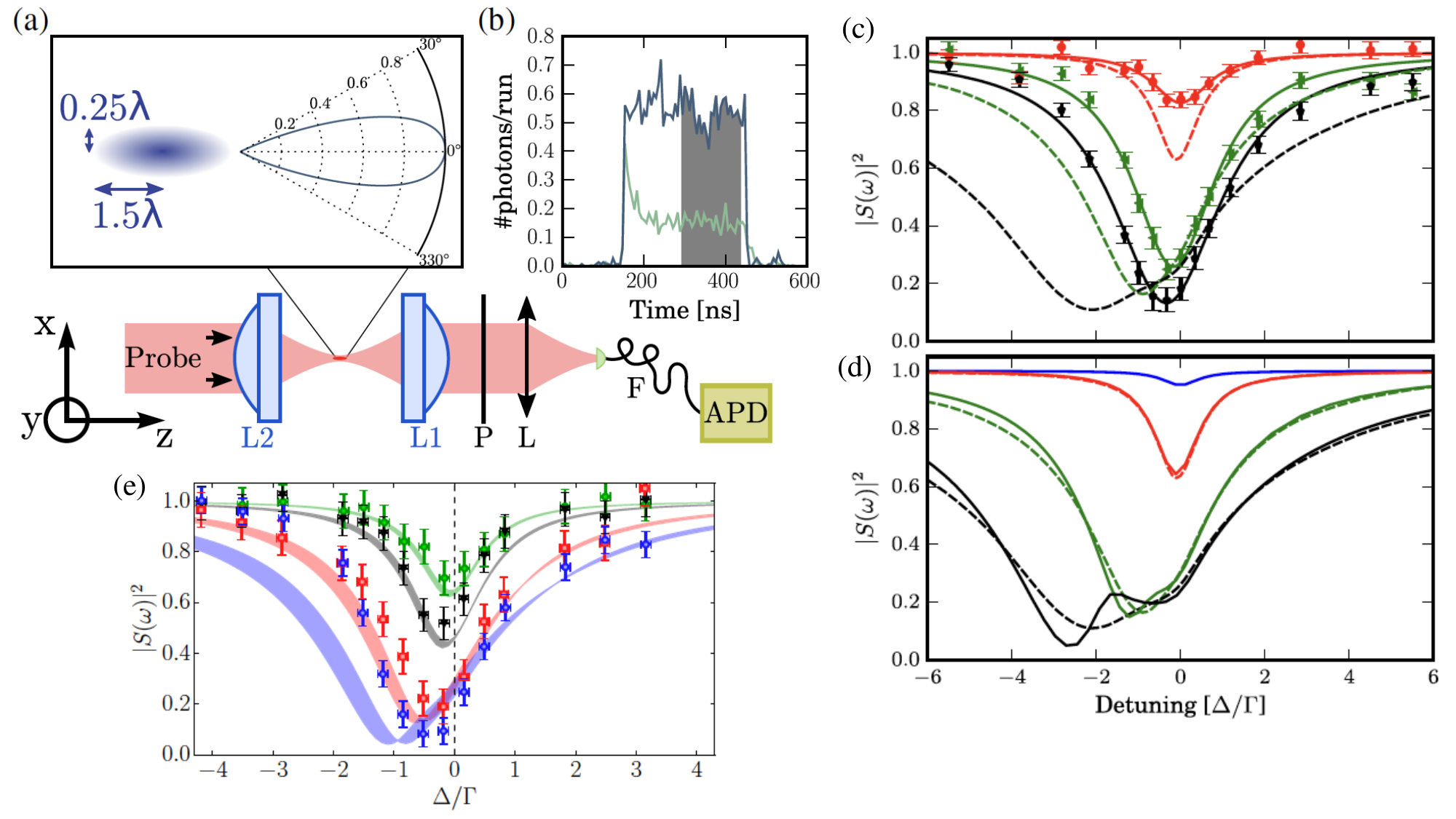}
	\caption{Coherent light scattering by a dense two-level atom cloud. (a) Experimental setup for a cloud of rubidium atoms illuminated by a linearly polarized probe laser. P is for polarizer, L for Lens, and APD for avalanche photo diode. Inset shows the cloud rms widths (left) and the intensity radiation pattern (right). (b) Example of temporal signals recorded on the APD with $N=180$ atoms (green line) and without atoms (blue line). Gray area: time interval used for the steady-state analysis. (c) Measured transfer function of the cloud in steady state versus probe detuning $\Delta$ for $N=(10$, $83$, $180)$ atoms (top to bottom); error bars: statistical (one standard deviation), shot noise limited. Solid lines: Lorentzian fit by $|S(\omega)|^2$. Dotted lines: results of the coupled dipole equations including the $12$ levels of the level transition. (d) Comparison between the predictions of the Lorentz model (solid line) and the microscopic model (dotted line) in (c) for $N=(1, 10, 83, 180)$ (top to bottom). Adapted from Ref. \cite{Jennewein2016}. (e) Transfer function as a function of the probe light detuning for various atom numbers. (Green, black, red, blue): $N = (10,20,60,100)$. The thick solid lines are the results of the multimode Maxwell-Bloch model with no adjustable parameters. The thickness of the lines reflects the uncertainties on the cloud sizes and on the waist of the probe laser. Adapted from Ref. \cite{Jennewein2018}.}\label{weak2}
\end{figure}

Under the assumption of weak excitations, there would be unavoidable mismatch between theory and experiments as seen in Fig. \ref{weak2}. This disagreement can be attributed to, from experimental perspectives to name a few, uncertain experimental parameters like atom cloud size, laser beam width, atomic ensemble temperature, number of atoms, and cloud geometry, which can significantly influence the effect of PMDDIs leading to superradiance  \cite{Rehler1971}. Meanwhile, most of the mean-field models neglect or keep up to certain degrees of high-order quantum correlations \cite{Kubo1962, Williamson2020, Robicheaux2021, Robicheaux2023, Bigorda2023, Kusmierek2023, Wang2024}, which is imperative or otherwise would face the hierarchy problems due to exponentially increasing size of Hilbert space, leading to incessant relations among different orders of correlations. This shows a conundrum of insolvability in almost all dynamical processes in open quantum systems with many-body interactions and as well the essence of atom-atom correlations indispensably induced in strong light-matter interacting systems, which cannot effortlessly included in most of the mean-field treatments. 

In Fig. \ref{weak2}, a pristine experiment of light scattering is conducted, and the transmitted light is measured and collected by a single-mode fiber in the steady state after a transient time around the transition lifetime $1/\Gamma$. In the low light limit, the coherent scattered field is observed with saturated extinction, an increasing small redshift, and a linewidth broadening as $N$ increases. This shows qualitatively the effect of PMDDIs, which deviates from the Lorentz-Lorenz formula in Fig. \ref{weak2}(d). Even for a microscopic dipole model as compared in Fig. \ref{weak2}(c), the predictions differ quantitatively and indicate the role of multiple scattering at all orders. In Fig. \ref{weak2}(e), a followup multimode and paraxial Maxwell-Bloch model is applied with agreements in low density regime of cold atoms, but disagrees with experiments at higher densities. Therefore, a quantitative understanding of PMDDIs in relatively simple setups still remains a challenge, and it demands more sophisticated theoretical approaches like quantum Langevin equations \cite{Scully1997, Jen2012, Jen2022_EIT}. 

\subsection{Finite excitations}

When light drives the system with finite excitations, weak field assumptions are put into question and a finite and limited Hilbert space for atomic excitations is no longer genuine. In terms of Dicke's formalism \cite{Dicke1954}, various spin sectors with respective fixed spin magnetization or atomic excitations can be populated, spanning the dynamical processes of light-matter interactions and light re-emissions into fully-connected state manifolds. Considering the superradiance from an inverted atomic clouds, a symmetric sector of pairwise PMDDIs can be invoked to truncate the dynamical spaces massively, and a quartic to quadratic factorization of atom-atom correlations further reduces the dynamical complexity into a manageable calculation more than a hundred of atoms \cite{Sutherland2017}.  

\begin{figure}[tb]
	\centering
	\includegraphics[width=0.85\textwidth]{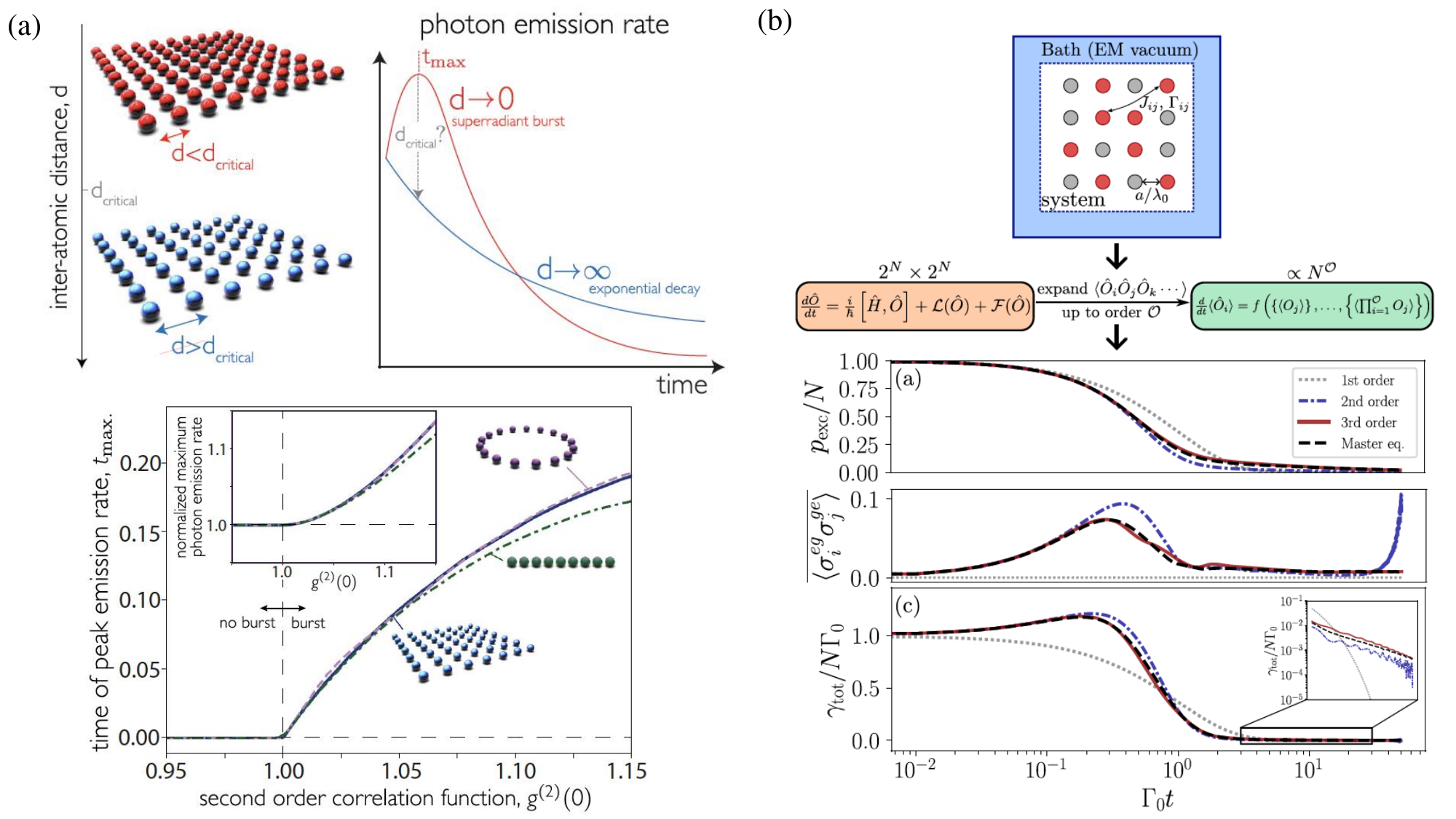}
	\caption{Many-body decay, cumulant expansion approach, and comparisons with different levels of cumulant expansions. (a) Inverted atomic array interact with each other and decay collectively via the burst light emission with a peak at time $t_{\rm max}$. This Dicke superradiance is contrasted with atoms far apart emitting exponentially as single entities, which can be distinguished by a critical interparticle distance. The bottom plot shows the second-order correlation function $g^{(2)}(0)$ at $t = 0$, where $t_{\rm max}>0$ only if $g^{(2)}(0)>1$. Inset: Maximum intensity, normalized by intensity at $t = 0$. In both plots, all nine atoms are initially excited, with polarization perpendicular to the array. Adapted from Ref. \cite{Masson2022}. (b) The vacuum mediates interactions among arrayed emitters and superradiant dynamics is characterized for fully or partially excited arrays with no coherences among emitters at $t = 0$. The time dynamics are compared via cumulant expansion of orders $1-3$ and with the solution of the full master equation for a chain of $N = 10$ atoms with lattice spacing $a = 0.1\lambda$. Three subplots are shown for the excited-state population, average pair correlations, and total emission rate. The inset in the bottom panel shows the total emission rate at late times in logarithmic scale. The atoms are considered to be polarized in the direction perpendicular to the lattice plane. Adapted from Ref. \cite{Bigorda2023}.}\label{inverted}
\end{figure}

Under finite light excitations, several recent progresses in experiments aim to study superradiant emissions from almost fully inverted atoms in free space \cite{Ferioli2021} and near waveguide \cite{Liedl2024, Bach2024, Tebbenjohanns2024}. The inverted systems illustrate superradiant burst \cite{Ferioli2021, Masson2020, Masson2022, Masson2024} despite of strong frequency shifts of PMDDIs and non-exponential decay due to collective effects as shown in Fig. \ref{inverted}(a). In particular, second-order correlations are highly correlated with delayed times at the maximal photon emission rates, marking the onset of significant influence of PMDDIs at a critical interparticle spacing \cite{Masson2022}. In Fig. \ref{inverted}(b), detailed comparisons are conducted between various orders of cumulant expansions \cite{Kubo1962}, presenting the essential roles of higher-order quantum correlations induced through the decay processes from inverted atomic systems \cite{Bigorda2023}. For smaller lattice spacings and large atom numbers, errors can be accumulated due to truncated orders of correlations, and the performance of calculations can be worsened at long times due to the buildup of more sophisticated atom-atom correlations from subradiant sectors of the system \cite{Bigorda2023, Bigorda2022}. 

Following the work of laser-driven Dicke superradiance \cite{Ferioli2021}, a non-equilibrium superradiant phase transition is observed between low and high driving fields \cite{Ferioli2023}. These results can be modeled by effective atom number contributing to the cooperative radiations, but a quantitative agreement on the measured second-order correlations is missing due to neglect of spatial correlations among extended atoms \cite{Ferioli2023}. A cluster expansion method \cite{Hazzard2014} is conducted to compare the experiments with various theoretical models with different degrees of involvement of atom-atom correlations \cite{Agarwal2024}. This method adopts a finite size of local clusters along with mean-field treatments on interactions with particles outside the clusters, which complements the mean-field model and bridges the gap between the mean-field (dipole method) and full treatments (exact diagonalization). A scaling of $N$ to $\sqrt{N}$ of the critical drive parameter is shown to mark the transition from superradiance to normal regime of cooperative resonance fluorescence \cite{Agarwal2024}. Although there are some fair agreements with experiments in Ref. \cite{Ferioli2023}, an open question still remains in discrepancies of the second-order correlation measurements at a moderate driving field.   

This mismatch between theories and experiments are mainly related to neglect or limited inclusion of atom-atom correlations, where even pristine experiments like light scattering measurements can nonetheless be endowed with high-order atom-atom \cite{Ferioli2023, Agarwal2024, Bach2024, Tebbenjohanns2024} or non-Gaussian correlations in dense atomic clouds \cite{Ferioli2024}. Some recent theoretical attempts to involve quantum fluctuations with truncated Wigner approximation \cite{Mink2022, Mink2023} are conducted as a semi-classical approach to account for quantum correlations among atoms induced by PMDDIs. This quantum noise model originates from the Fokker-Planck equation \cite{Scully1997} used in describing fluctuations in Brownian motion \cite{Gardiner2004}, which is equivalent to quantum Heisenberg-Langevin equation and has been applied in laser theory \cite{Haken1970} in the positive-$P$ phase representation \cite{Gardiner2000, Drummond1980, Drummond1987, Drummond1991, Jen2012, Kiesewetter2023}. Quantum Langevin equations with Gaussian quantum fluctuations can involve up to the second-order quantum noise correlations with a self-consistent account of quantum noises in atomic and field variables. Meanwhile, signatures of non-Gaussian correlations \cite{Ferioli2024, Horvath2024} indicate higher-order quantum correlations among the atoms, where a concrete and applicable theoretical model of non-Gaussian baths for dense atomic clouds still remains elusive \cite{Funo2024}. The main reason is that it is challenging to treat quantum mechanically with a full-scale of light and atomic degrees of freedom in light-matter interacting systems, simply due to the exponentially growing joint quantum states accessible in dynamically coupled processes. 

As a final remark, PMDDIs can lead to several potential applications in quantum technology, such as precision spectroscopy using anti-resonance of subradiant arrays in cavities \cite{Plankensteiner2017}, collective enhanced cooling \cite{Vogt1996, Xu2016, Maximo2018, Gisbert2019, Wang2023, Bigorda2024}, and steady-state superradiant lasers \cite{Haake1993, Meiser2009, Meiser2010, Bohnet2012, Maier2014, Ludlow2015, Jen2016_SL}. The superior cooling performance can be achieved via subradiant collective resonances with narrowed linewidths \cite{Bigorda2024}, while the dominance of atomic coherences in the bad cavity limit leads to a synchronization of dipoles in a driven atom-cavity system, promising an extremely narrow linewidth \cite{Meiser2009} and enabling a continuous and collective emission for next-generation atomic clocks. For fundamental scientific studies, PMDDIs are also shown to provide a cooperative mechanism for photon localization \cite{Akkermans2008, Bienaime2012}. This can be leveraged to find applications in quantum storage of light or quantum information processing \cite{Hammerer2010}.   
 
 \begin{figure}[tb]
 	\centering
 	\includegraphics[width=0.8\textwidth]{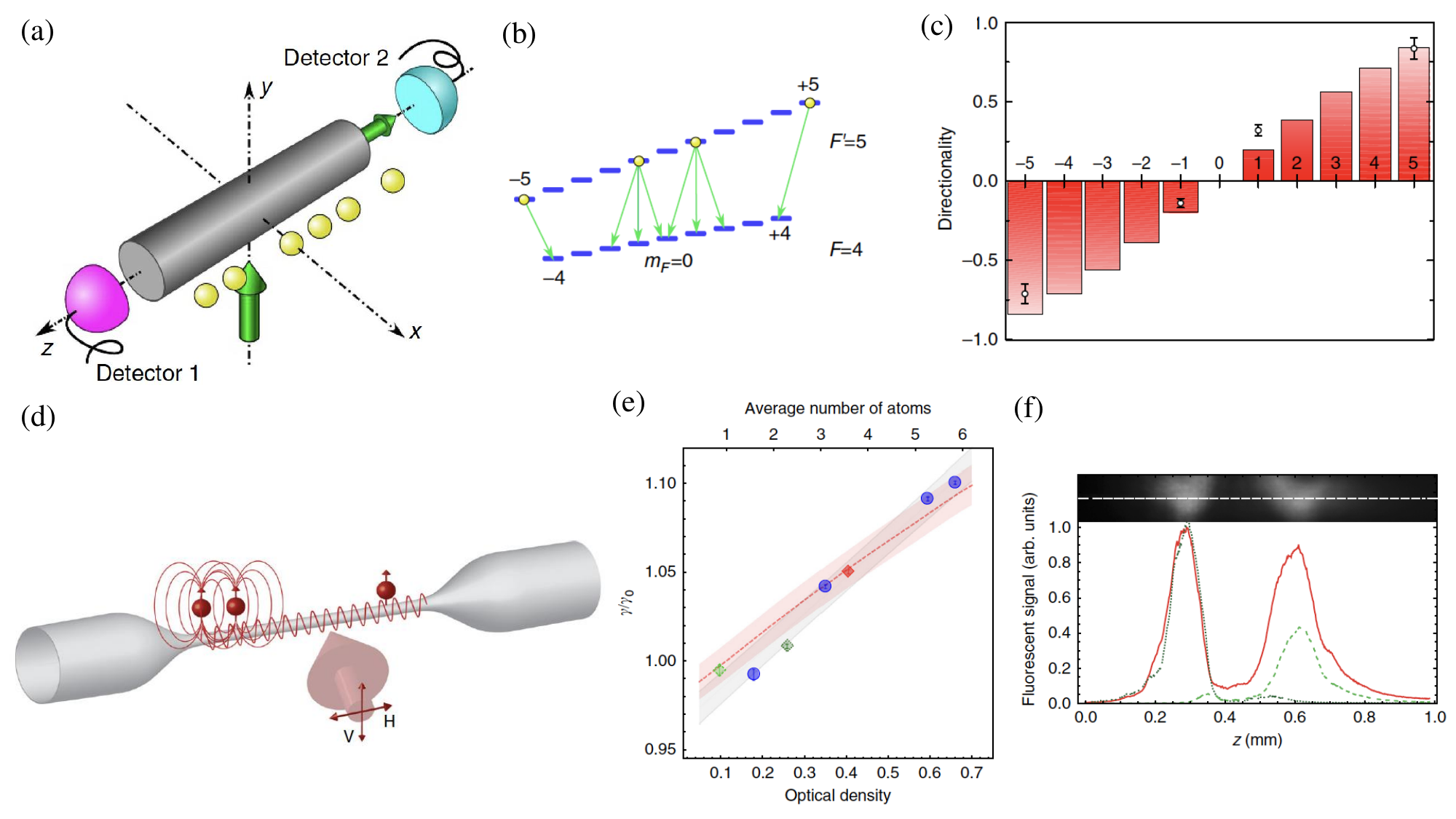}
 	\caption{Schematic plots of experiments with quantum state-controlled directional spontaneous emission \cite{Mitsch2014} and superradiant decay among two separated atom clouds \cite{Solano2017}. (a) Experimental setup with atoms (yellow spheres) trapped on one side of the nanofiber. A left-handed circularly polarized laser beam propagating in the $y$-direction (vertical green arrow) excites the atoms. The fluorescence light emitted from the atoms into the nanofiber is recorded by two detectors at each end of the waveguide. (b) Atomic level scheme with the initially excited atomic states (yellow spheres) and the decay channels (green arrows) controls the amount of decay for different light polarizations. (c) Directionality of the spontaneous emission into the waveguide as a dependence of the magnetic quantum number of the excited state. Open circles: averaged measurement results. The error bars correspond to the deviation estimation and are compared with theory predictions denoted by bars in good agreement. Adapted from Ref. \cite{Mitsch2014}. (d) Schematic of a platform for collective radiations from singly excited atomic states. When two atoms are close to each other, the dipolar interaction is mostly mediated by the electromagnetic radiation modes outside the nanofiber, which decays inversely with distance, while widely separated atom clouds are mediated by the guided mode on the waveguide for arbitrary distances. (e) The normalized fast decay rates as a function of the optical density and N. The blue circles correspond to the signals from a single cloud of atoms. The atomic cloud is splitted in two as shown in (f).The dashed light and dotted dark green diamonds, and the solid red square correspond to the right, left, and the combination of both atomic clouds, respectively. The gray region represents the one-sigma confidence band of a linear fit to the data, the red dashed line is the theoretical prediction, and the red shaded region represents a confidence interval set by a fractional error of $1\%$. (f) Separated atom clouds show long-range interactions, where the top of the figure shows in black and white a fluorescence image of a split cloud. The white dotted line represents the waveguide location. The fluorescence signal of the split clouds along the nanofiber is plotted as a function of position. The dashed light (dotted dark) green dashed lines is the intensity distribution of the right (left) atomic cloud when the other one is blocked. The solid red line is the intensity distribution when both clouds are present. The separation between the center of both clouds is around 318 $\mu$m, equivalent to around $400$ wavelengths. Adapted from Ref. \cite{Solano2017}.}\label{WQED}
 \end{figure}
 
\section{Waveguide QED}

Light can be strongly confined in the waveguide such as photonic crystal waveguide or nanofiber and leads to enhanced light-atom couplings when respectively the atoms are planted in these nanostructures like quantum dots \cite{Sollner2015, Young2015} or put close to these photonic nanostructures like neutral atoms trapped in optical lattices \cite{Corzo2019, Sheremet2023, Vetsch2010, Johnson2019}. In such 1D atom-waveguide platforms \cite{Douglas2015, Tudela2015, Chang2018} as illustrated in Fig. \ref{WQED}, chiral quantum optics \cite{Lodahl2017} can emerge due to the spin-momentum locking via evanescent waves \cite{Bliokh2014, Bliokh2015} on the surface of these waveguides, breaking the time-reversal symmetry that is often conserved in conventional free-space quantum optical platforms. This offers many potentially new research directions like modified collective radiations from non-reciprocally coupled atoms, optical manipulations of quantum information, and simulations of exotic many-body states \cite{Lodahl2017}.  

Here we focus on the waveguide QED systems with PMDDIs \cite{Kien2005, Kien2008, Chang2012, Loo2013, Goban2015, Shahmoon2016, Ruostekoski2016, Ruostekoski2017, Kien2017, Solano2017, Pichler2015}. The interaction form presents infinitely long-range spin-exchange interactions \cite{Solano2017} as shown in Eq. (\ref{chiral1D}) and can even lead to chiral \cite{Gardiner1993, Carmichael1993, Stannigel2012, Downing2020} or nonreciprocal couplings among the atoms \cite{Lodahl2017, Mitsch2014, Pichler2015}. This promises new platforms of quantum interface to manipulate light-matter interactions and achieve the strong coupling regime that is difficult for atoms in free space. Below I review two main developments on photonic transport and quantum simulations, which spur studies for fundamental quantum science and applications in quantum technology. There are atom-nanophotonic cavity interfaces that can host strong couplings \cite{Samutpraphoot2020, Liu2023, Li2023} and demonstrate the potential for scalable entanglement transport \cite{Dordevic2021} and graph state generations \cite{Cooper2024, Chien2024}. With chiral couplings in such interfaces, W states can be generated \cite{Hiew2023}, useful for quantum information processing.  

\subsection{Photon transport}

Due to the long-range nature of PMDDIs in the atom-waveguide systems, numerous opportunities emerge for studying photon-photon correlations and for applications in routing optical information via distinct manipulations of light and matter.  In Fig. \ref{WQED}, two major demonstrations showcase the essence of infinite-range PMDDIs and tunable directionality of couplings to the waveguide. The former presents the all-to-all couplings among atoms showing the ultimate longest-range as well as strongly interacting systems via the guided modes supported on the waveguides, while the latter offers high control and flexibility in the ratios of coupled radiations between the left and the right propagating modes.   

A factor of $\beta\equiv (\gamma_{\rm L}+\gamma_{\rm R})/\Gamma_{\rm tot}$ is useful to quantify the strength of coupling to the waveguide \cite{Lodahl2017}, where the total decay rate $\Gamma_{\rm tot}=\gamma_{\rm L}+\gamma_{\rm R}+\gamma_{\rm ng}$ is composed of the left (L)- and the right (r)-propagating modes, and non-guided modes $\gamma_{\rm ng}$. A $\beta\sim 1$ indicates the strong coupling regime where almost all emitted light scattered from atoms is coupled to the waveguide. Meanwhile, $\beta\ll 1$ denotes a weak coupling regime, where most of the light scattered to free space, similar to the single-particle scenario with its lifetime of dynamics purely determined by $\gamma_{\rm ng}^{-1}$. Generally speaking, $\beta$ can be as high as $0.99$ for quantum dots and $0.999$ for superconducting transmon qubits, but $N$ is limited to $\mathcal{O}(1)$ and $\mathcal{O}(10)$, respectively \cite{Sheremet2023}. On the contrary, atom-waveguide systems can host $10^3$ neutral atoms, but the coupling efficiency can only reach $\beta\sim 10^{-2}$ \cite{Corzo2019}. This shows respective advantages for different platforms, and there is still plenty of room for improvement in reaching the strong coupling regime along with large number of coupled quantum emitters.
 
Interestingly, even for $\beta$ as low as $10^{-1}$, a universal strong bunching effect of photon-photon correlations can emerge along with a power-law decay of transmitted light power versus $N$, presenting a signature of nonlinear couplings in the quantum interface \cite{Mahmoodian2018}. Long-time decay behavior can also be hosted as subradiant states \cite{Henriet2019, Kornovan2019, Ke2019, Jen2020_subradiance, Kumlin2020, Berman2020, Pivovarov2021, Jen2021_bound} when $\beta$ is close to one to promise lossless energy transport and light storage \cite{Albrecht2019}. In a chiral quantum interface in the strong coupling regime, many-body photon bound states can be realized \cite{Zhang2020_bound, Mahmoodian2020}, photon-mediated localization can be induced by interactions \cite{Zhong2020}, and the absence of atomic excitation transport can be allowed under strong disorders \cite{Mirza2017, Jen2020_disorder, Jen2021_crossover, Jen2022_correlation}. Nonlinear optical effects of polariton scattering \cite{Schrinski2022} has been used in revealing the Lieb–Liniger model of the so-called Tonks–Girardeau gas \cite{Tonks1936, Girardeau1960} in the limit of many emitters, while photon-photon repulsion can emerge by controlling three-level quantum emitters, leading to self-ordering of regular trains of single photons \cite{Iversen2022}. Multiphoton transport in 1D waveguide interacting with multiple atoms is also theoretically treated to calculate the reflection and transmission properties \cite{Liao2020}.  

As for collective radiations in the atom-waveguide systems, recent progresses show superradiant decays from emissions in the forward-propagating guided mode along with increasing collective response of the atomic ensemble in a design of $45$-m-long fiber ring resonator \cite{Pennetta2022} and observations of subradiant states leading to sudden and temporary switch-offs of the emitted optical power \cite{Pennetta2022_2}. In the discussion of subradiant sectors and their scaling behaviors on $N$, recent theoretical investigations uncover that the scaling of most subradiant decay rates depends on the dispersion relation near the band edge of the infinite 1D system \cite{Zhang2020, Zhang2022}. A superradiant burst of light in a chiral waveguide can show up when above a threshold number of atoms with its peak power scaling even faster on the number of atoms than Dicke superradiance in free-space \cite{Liedl2024}. Lastly, a stochastic simulation based on the truncated Wigner approximation is applied to predict the second-order correlation function in such chiral waveguide with large $N$, but it has limitations in capturing the long-time system dynamics \cite{Tebbenjohanns2024}.  

Similar to the PMDDI-assisted cooling for neutral atoms in free space \cite{Wang2023, Bigorda2024}, a waveguide mediated PMDDI can also be utilized to showcase an imbalanced heat transfer among the target and residual atoms under nonreciprocal decay channels \cite{Chen2023}. This promises an enhanced performance of cooling rate in dark-state sideband cooling approach under asymmetric laser driving fields and illustrates new opportunities with PMDDIs in distinct heat transfer \cite{Wang2022}. PMDDI-assisted cooling can further overcome the cooling bottleneck as one of the limitations in scalable quantum computations \cite{Pino2021, Leibfried2003}.  
 
\begin{figure}[tb]
	\centering
	\includegraphics[width=0.85\textwidth]{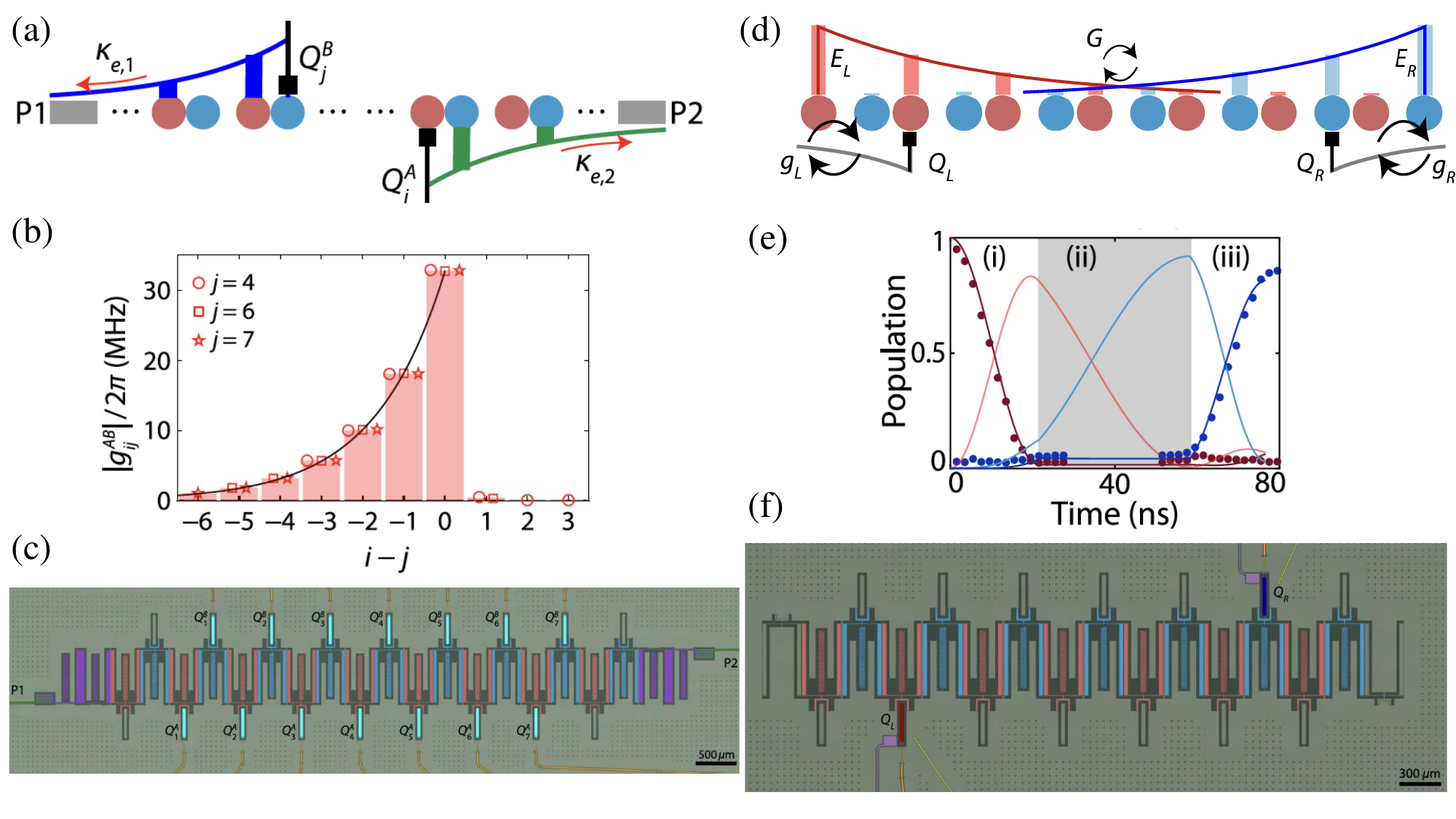}
	\caption{Directionality of qubit-photon bound states and topological edge modes. (a) Theoretical photonic envelope of the directional qubit-photon bound states for the device I, consisting of $9$ unit cells with qubits (black lines terminated with a square) coupled to every site on the $7$ central unit cells. At a reference frequency, the qubit coupled to site A (B) induces a photonic envelope to the right (left), colored in green (blue). (b) A plot of coupling between various qubit pairs from the experiments with a solid black curve in an exponential fit to the measured qubit-qubit coupling rate versus qubit index difference (spatial separation). (c) Optical micrograph of device I. Qubits (cyan, labeled as $Q^\alpha_j$ are coupled to every site of the $7$ inner unit cells of the topological waveguide. (d) Schematic of device II, consisting of $7$ unit cells. $E_L$ and $E_R$ are the left-localized and right-localized edge modes which interact with each other at rate $G$ due to their overlap in the center of the waveguide. (e) Population transfer from $Q_L$ to $Q_R$ with three consecutive swap transfers $Q_L\rightarrow E_L\rightarrow E_R\rightarrow Q_R$. The population of $Q_L$ ($Q_R$) during the process is colored dark red(dark blue), with filled circles and solid lines showing the measured data and fit from theory, respectively. The light red (light blue) curve indicates the expected population in $E_L$ ($E_R$) mode, calculated from theory. (f) A device II topological waveguide with $7$ unit cells [resonators corresponding to A (B) sublattice colored red (blue)]. Two qubits $Q_L$ (dark red) and $Q_R$ (dark blue) are coupled to A sublattice of the second unit cell and B sublattice of the sixth unit cell, respectively. Adapted from Ref. \cite{Kim2021}.}\label{TWQED}
\end{figure}

As a final remark in this subsection, we note of a topological waveguide \cite{Barik2018, Bello2019} which can be manipulated to show collective quantum dynamics and edge modes \cite{Kim2021}. In Fig. \ref{TWQED}, two devices of superconducting quantum circuits \cite{Krantz2019} can be fabricated to demonstrate directional qubit-photon bound states \cite{Sundaresan2019, Shi2018} and to perform quantum state transfer via topological edge states. With multiple quantum emitters of transmon qubits coupled to a topological waveguide, a photonic analog of the Su-Schrieffer-Heeger model \cite{Su1979}  and collective radiations can be exhibited. This paves the way toward quantum simulation applications in forming exotic quantum matter of light \cite{Douglas2015, Chang2018} and synthesizing long-range interacting quantum spin models \cite{Hung2016}. In addition to this potential quantum simulation framework using novel types of photon-mediated interactions in topological waveguide QED \cite{Kim2021}, below we further explore driven-dissipative setups that are particularly of interests in simulating non-equilibrium quantum dynamics in atom-waveguide interfaces. 

\subsection{Quantum simulation of many-body quantum states in driven-dissipative setup}

In open quantum systems \cite{Breuer2002, Scully1997}, driven-dissipative setups can be appropriately modeled by adding a coherent driving term $H_{\rm d}=\sum_{j=1}^N\Omega\hat\sigma^\dag_\mu\hat\sigma_\mu$ with a Rabi frequency $\Omega$ to the Hamiltonian in Sec. \ref{Theory}. Along with the Liouvillian $\mathcal{L}$ in Lindblad form for dissipative dynamics of atoms, the whole system can be analyzed and resolved by solving its master equation in time. In particular for the steady-state solutions, the system can be dissipatively evolved into unique and desired many-body entangled states \cite{Diehl2008, Kraus2008, Verstraete2009} by reservoir engineering. For transient time dynamics, new quantum phases of matter can undergo non-equilibrium phase transitions driven by dissipation and interaction \cite{Diehl2010}. Furthermore, robust edge states and non-Abelian excitations as topological states of matter, and corresponding topological protection can also be demonstrated in open quantum systems with engineered dissipation \cite{Diehl2011}.   

\begin{figure}[tb]
	\centering
	\includegraphics[width=0.9\textwidth]{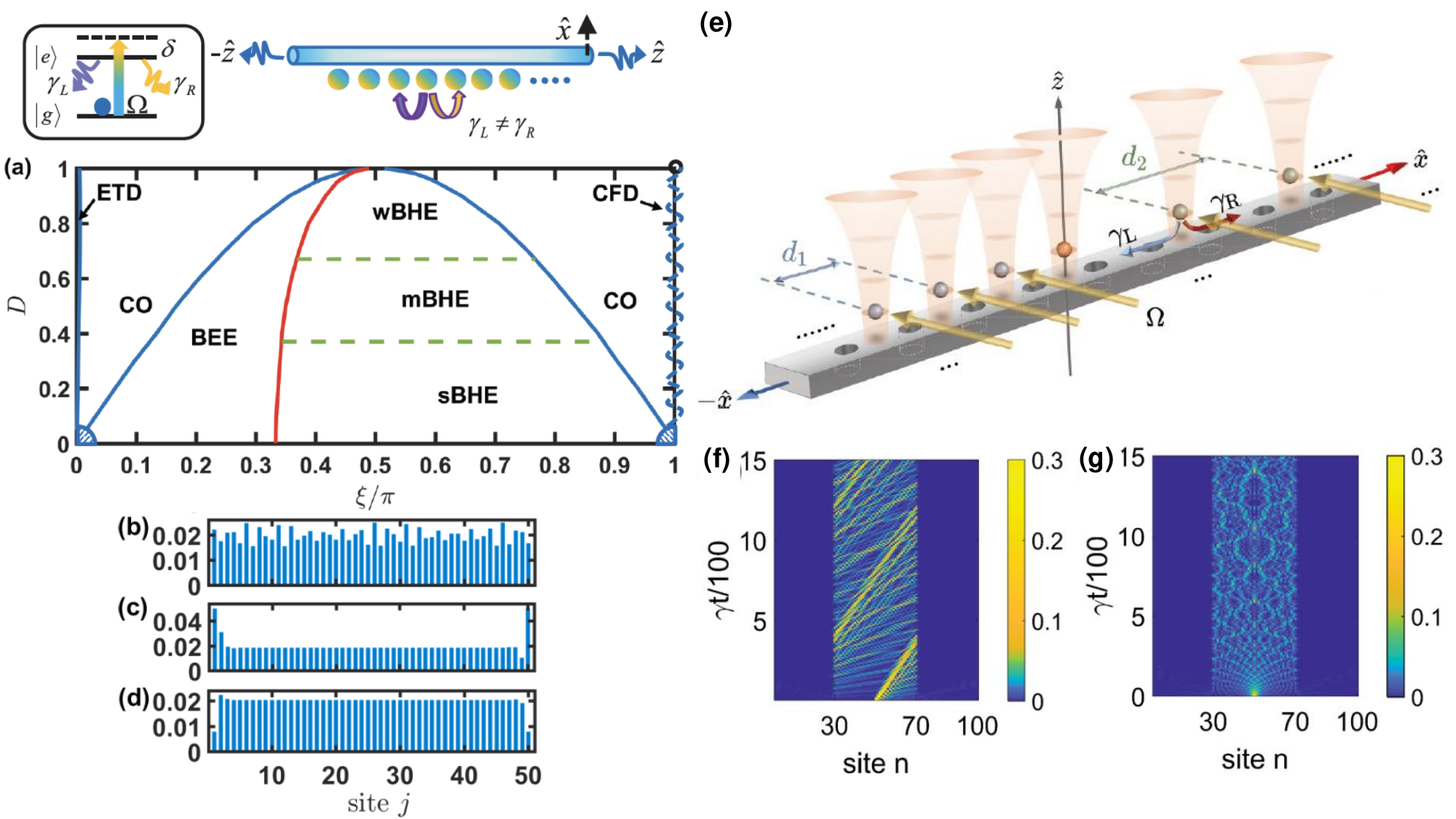}
	\caption{Steady-state phase diagram of a weakly driven chiral-coupled uniform and dissimilar atomic chain. (a) The steady-state phases involve extended distributions (ETDs), finite crystalline orders (COs), bi-edge/hole excitations (BEEs/BHEs), and a region of chiral-flow dichotomy (CFD, wavy dashes), which are identified under parameter spaces of directionality factor $D\equiv (\gamma_{R}-\gamma_{L})/(\gamma_{R}+\gamma_{L})$ \cite{Mitsch2014} and dimensionless interatomic distance $\xi$. The BHE phase is further separated into strong (s) (below $D\approx 0.37$), moderate (m), and weak (w) regimes (above $D\approx 0.67$). The shaded regions at the two lower corners of the diagram represent the critical regimes with divergent population distributions. (b,c,d) CO phase at $D = 0.2$, $\xi/\pi=0.02$, BEE phase at $D = 0.2$, $\xi/\pi=0.2$, and sBHE phase at $D = 0.2$, $\xi/\pi=0.6$, respectively, for $N=50$. Adapted from Ref. \cite{Jen2020_steady}. (e) A schematic of weakly driven dissimilar chirally coupled dissimilar atomic array coupled to photonic crystal waveguide. Adapted from Ref. \cite{Chung2024}. (f, g) Excitation trapping effect in dissimilar arrays. The atomic excitation can be confined by introducing two interfaces with a design of three zones with $\xi_{1,2,3}=\pi/2,~\pi,~\pi/2$ (left, middle, and right zones) for a total $N = 100$ with $N_{1,2,3}=30, ~40, ~30$, respectively, where two interfaces are at the sites $n = 30$ and $70$. The excited-state populations is plotted for (f) $D = 0.2$ and (g) $D = 0$ under a single-site excitation in the center of the middle zone. In (g), we study the case at $\xi_2=\pi/8$ instead. Adapted from Ref. \cite{Handayana2024}.}\label{SS}
\end{figure}

Similarly in atom-waveguide interfaces, an entangled mesoscopic steady state can be collectively prepared by controlling the qubit positions with assistance of a coherent driving and engineered spontaneous decay \cite{Tudela2013, Buonaiuto2019}. Collective polaritonic modes can be formed \cite{Kornovan2016}, and scattering of a laser field into the waveguide modes can be dramatically enhanced away from the geometric Bragg angle \cite{Birkl1995, Schilke2011} due to waveguide-mediated atom-atom interactions \cite{Olmos2021}. Under weakly driven-dissipative conditions as shown in Fig. \ref{SS}(a), the steady-state phases with crystalline orders or edge excitations can show up due to the interplay between PMDDIs and directionality of couplings \cite{Jen2020_steady}. A nonergodic butterfly-like system dynamics at long time with agglomeration of atomic excitations around the array center is demonstrated with a signature of persistent subharmonic oscillations \cite{Jen2020_steady}, resembling the quantum many-body scars in the transverse field quantum Ising model on a ladder \cite{Voorden2020} or driven Rydberg chains as spin-1/2 XX Heisenberg chain \cite{Voorden2021}. The so-called scar states \cite{Turner2018_1, Turner2018_2, Lin2019} are associated with confined populations in some subset of the excited states, originally discussed in isolated and unstable periodic orbits in a classical chaotic system \cite{Heller1984}, which break the ergodicity of the system. These distinct features of nonequilibrium dynamics can be traced to quantum Newton's cradle that breaks ergodicity \cite{Kinoshita2006}, related to the thermal equilibrating process in quantum chaotic systems \cite{Tsaur1998, Srednicki1999}. The nonergodic phenomena are similar to persistent oscillations in Rydberg atoms \cite{Bernien2017} or discrete time-crystalline orders in many dipolar spins in a diamond \cite{Choi2017}. 

When an ordered atomic array in the atom-waveguide coupled system is concatenated with another array with different interparticle spacings, a dissimilar atomic array setup can be formed as shown in Fig. \ref{SS}(e). This presents fruitful opportunities to engineer system dynamics and tailor exotic quantum states. With three zones of atomic arrays at the chosen parameters in Figs. \ref{SS}(f) and \ref{SS}(g), excitation trapping can be formed and long-time interference patterns emerge in quench dynamics \cite{Handayana2024} due to the spin-exchange couplings of PMDDIs. Lastly, atom-waveguide interfaces can be used to study atomic excitation delocalization behaviors \cite{Anderson1958, Clement2005, Evers2008} at the clean to disordered interface \cite{Wu2024}. A design of two zones can provide insights to explore the influence of the baths on excitation localizations. This opens new directions in unraveling the fate of many-body localization \cite{Luitz2017, Roeck2017, Thiery2018, Morningstar2022, Sels2022, Leonard2023} in open quantum systems \cite{Luschen2017, Fayard2021} and promises novel applications in quantum state engineering \cite{Chung2024}. 

\section{Topological quantum optics in cold atoms} 

\begin{figure}[tb]
	\centering
	\includegraphics[width=0.95\textwidth]{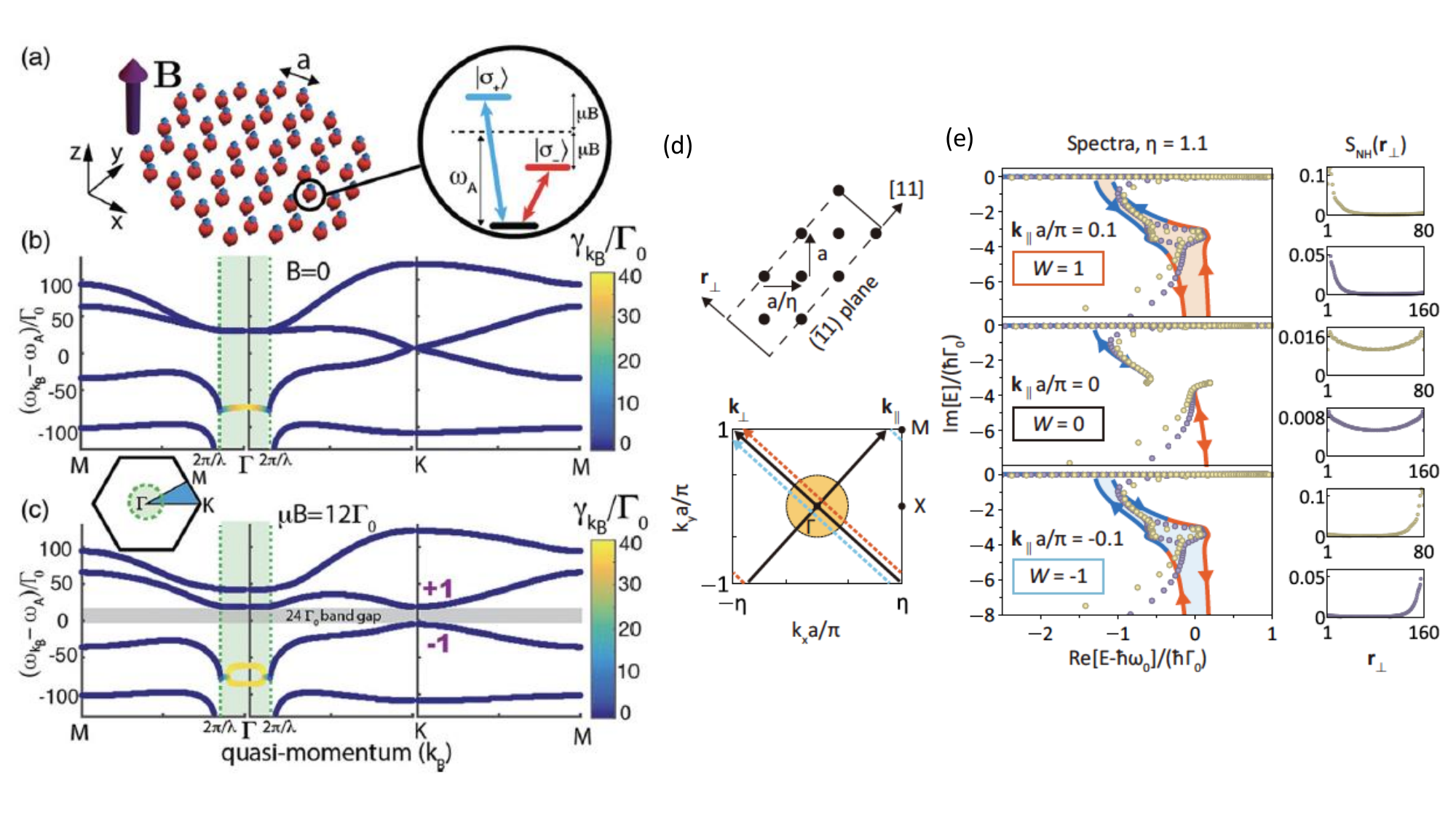}
	\caption{Complex energy band structure and geometry-dependent NHSE in a ribbon geometry. (a) Honeycomb lattice of three-level V-type atomic emitters with interatomic spacing $a$. A magnetic field breaks the degeneracy via the Zeeman splitting. (b) Band structure of the lattice with vanishing $B$. Green dashed lines indicate the free-space light cone and quasi-momentum modes $k_B < \omega_{k_B}/c$ couple to free-space modes with enhanced decay rates (green shaded region). Decay rates of the modes are color coded. Bands are degenerate at the symmetry points \textbf{K} and $\bf \Gamma$. (c) A transverse magnetic field opens a gap (grey shaded region) between topological bands with nontrivial Chern numbers. Relevant parameters are $\lambda=790$nm, $\Gamma/(2\pi)=6$MHz, and $a=0.05\lambda$. Adapted from Ref. \cite{Perczel2017}. (d) Illustration of the ribbon geometry for a rectangular atomic lattice. The boundaries are open on the~$(\bar{1}1)$ plane~(dashed line) and extend infinitely in the~$[11]$ direction~(solid line). Open boundary eigenenergy spectra of rectangular~(lattice constant ratio $\eta=1.1$) lattices in ribbon geometries with a width of $80$ and $160$ unit cells~(light yellow dots and purple dots, respectively) and the corresponding bulk spectra~(curves in red and blue) at fixed $\mathbf{k}_{\parallel}=0.1\pi/a$~[top panels, orange dashed line in (d)] and $\mathbf{k}_{\parallel}=0$~[middle panels, $\mathbf{k}_{\perp}$ axis in (d)] and $\mathbf{k}_{\parallel}=-0.1\pi/a$~[bottom panels, cyan dashed line in (d)]. The non-Hermitian parts of effective energy spectra result in the nontrivial winding~[top and bottom panels in (e)], and the corresponding spatial distributions~(right columns) demonstrate the extensive skin modes localized at the edge normal to $\mathbf{r}_{\perp}$ axis in a ribbon geometry. Adapted from Ref. \cite{Wang2022_skin}.}\label{Topology}
\end{figure}

Here we switch gears to topological quantum optics, in contrast to conventional quantum optics that does not generally manifest topological properties. Topology, originated from a mathematical object, indicates a robust physical property which can sustain through continuous deformations \cite{Cooper2019}. In addition to the topological properties tailored in photonics \cite{Ozawa2019} or ultracold atoms \cite{Cooper2019}, PMDDIs in neutral atomic arrays \cite{Shahmoon2017, Rui2020} provide another mechanism to engineer topological states, even under significant losses in light scattering in the optical domain \cite{Perczel2017, Perczel2017_2, Perczel2020}. In the quasi-momentum spaces for PMDDIs in a dense 2D atomic array under single excitation eigen-analyses, an infinite periodic honeycomb lattice has shown confined and subradiant modes outside the light cone, where an external magnetic field induces atomic Zeeman shifts and creates a band gap in the optical excitation spectrum with nontrivial Chern numbers and topologically protected edge states \cite{Perczel2017}, as shown in Fig. \ref{Topology}(a-c). This further spurs many new directions in controlling interactions between impurity atoms within an atom array \cite{Patti2021}, optimized photon storage \cite{Buckley-Bonanno2022}, topological transitions in atomic arrays coupled to a cavity waveguide \cite{Mann2022}, creation of topological photonic states in the complex Zak phase \cite{Wang2018}, and emergence of quantum Hall phases and topological edge states due to polariton-polariton interactions in the two-particle Hilbert space of an atomic array coupled to a waveguide \cite{Poshakinskiy2021}. 

In addition to these quasi-momentum confined modes outside the light cone, which showcase intriguing topological properties in the optical domain, the non-Hermitian modes with finite dissipations inside the light cone also manifest rich phenomenon that has no Hermitian counterparts. This relates to the non-Hermitian realm of quantum systems \cite{Ashida2020, El-Ganainy2018}, where two central phenomena of exceptional points (EPs) \cite{Bergholtz2021} and non-Hermitian skin effect (NHSE) \cite{Lee2016, Yao2018, MartinezAlvarez2018, Kunst2018, Lee2019, Borgnia2020, Okuma2020, Zhang2020_skin, Wang2024_skin} emerge. The former results from a coalesce of both complex eigenvalues and eigenstates of a non-Hermitian system, while the latter represents an extensive number of localized eigenstates at the boundaries under open boundary conditions. The EPs can emerge from a Hermitian system with Dirac or Weyl points along with added non-Hermiticity from dissipations and results in nontrivial windings in the complex energy plane. This leads to NHSE in one-dimensional systems and may illustrate geometry-dependent NHSE in higher-dimensional metasurfaces \cite{Zhang2022_skin, Wang2022_skin}. 

The geometry-dependent NHSE is studied in detail in a 2D atomic array with PMDDIs and with parallelogram-shaped boundaries, which relies on the orientations of open boundaries (oblique angles) and the lattice configurations (broken mirror symmetry). This can also be evidenced in a ribbon geometry with a bulk energy spectra enclosing nonzero spectral areas \cite{Wang2022_skin} in Fig. \ref{Topology}(d-e). Similar to the scale-free non-Hermitian skin modes predicted in 2D atomic arrays, a long-range coupled Hatano-Nelson model \cite{Hatano1996} also gives rise to the scale-free skin modes \cite{Wang2023_scaling}, offering a understanding on the interplay between long-range couplings and non-Hermiticity. NHSE is further investigated in a non-Hermitian fermionic chain under a quasiperiodic potential \cite{Wang2023_skin} to show its influence on the thermal and many-body localized phases \cite{Abanin2019, Hamazaki2019}. 

Non-Hermitian quantum metasurfaces have been demonstrated in diverse physical platforms, including photonic crystals \cite{Zhen2015, Zhou2018}, topolectrical circuits \cite{Lee2018, Hofmann2020}, and collectively dissipative cold atoms \cite{Wang2024_nexus}. These light-matter interacting platforms host not only topological excitations, but also optical nonlinearities that can demonstrate new collective behaviors. An interplay between non-Hermiticity, topology, and long-range interactions can provide unique opportunities to control quantum properties of light and matter, enhance the optical response, and promise novel quantum devices in creating exotic multiply-entangled states by tailoring PMDDIs in an atomic array \cite{Wang2022_skin}. Finally, the quantum metasurfaces of atomic arrays can find potential applications in topological quantum computing, quantum network, and quantum simulation of strongly correlated states. 

\section{Summary and Outlook} 

In this review, we present the essential progresses and ongoing research directions manifesting PMDDIs as a resource for quantum science and quantum technology. We have introduced the main theoretical approaches that provide many distinct predictions showcasing the capability of collective atom-atom interactions. Cooperative light scattering from atom clouds in free space, collective radiations in a waveguide QED system, and topological quantum optical platforms are reviewed in detail, which promise quantum engineering of exotic many-body states, novel applications in light manipulations, and enhanced performance in quantum metrology and quantum computation.    

There are difficulties and even conundrums encountered in both theoretical treatments and experimental interpretations, which nonetheless stimulates further studies in designing new theoretical approaches and future experiments for clarification and better agreements. One essential consideration is the role of quantum correlations induced, emerged, and evolved throughout the light-atom interactions, which significantly modifies the outcome of light scattering spectrum and its second-order correlations \cite{Ferioli2024}. New attempt of confining the involved Hilbert spaces under symmetry \cite{Holzinger2024_2} can sufficiently reduce the cost of computations in solving the dynamics of superradiance and provide insights to early-time many-body dynamics. Stochastic methods with controlled orders of correlations in light scattering \cite{Tebbenjohanns2024, Mink2022, Mink2023} setups can also offer a limited but reasonable inclusion of the effect of finite high-order correlations for converging and genuine predictions of system dynamics \cite{Kusmierek2023}. What is challenging lies at the finite excitations and the collective radiations that follow, where mean-field models are not sufficient in describing high-order correlations observed in experiments \cite{Ferioli2023}, especially in dense atom clouds with significant PMDDIs. 

Apart from light scattering in free space, quantum metasurfaces with subwavelength atomic arrays or meta-atoms of plasmonic nanoparticles have raised paramount interests in the quantum information and quantum technology communities. With collective and PMDDIs in arrays of atoms, quantum metasurfaces have shown rich opportunities in light manipulations, generations of many-body entangled photonic states \cite{Bekenstein2020}, and realizations of exotic topological phases \cite{Perczel2017} when coupling them to nanoscopic waveguides \cite{Douglas2015, Chang2018}. New directions of embedding a structured 2D atomic array near one-sided cavity or inside a cavity can enable reservoir engineer to tailor the PMDDIs among the atoms, for example, by tuning cavity widths \cite{Mann2022}. This leads to nontrivial quasi-momentum band engineering, distinct topological properties, and exotic collective radiation behaviors. These intriguing phenomena result from the interplay between non-Hermiticity of the system, topology, and long-range interactions mediated by light. In addition to 1D waveguide QED systems \cite{Sheremet2023}, new strongly coupled regimes between light and atoms can emerge in 2D waveguide QED \cite{Marques2021, Tecer2024}, which has unique long-rang PMDDIs with extending distances in between the 1D (infinite-range \cite{Solano2017}) and 3D cases (resonant dipole-dipole interactions in free space \cite{Lehmberg1970}) of reservoirs. This can lead to long-lived two-photon repulsive and bound states, and photon-photon correlations. 

There are several new opportunities to turn PMDDIs into useful resource for quantum metrology or novel quantum devices. In engineered arrays of narrow-band cubic lattices of strontium atoms \cite{Hutson2024}, systematically resolving their cooperative frequency shift in the millihertz level from PMDDIs can further boost the precision of optical clocks. It also provides a new avenue to explore many-body dynamics in atomic arrays based on controlled collective light-atom interactions. PMDDIs can also lead to recoiled momentum on the atoms, forming an alternative but distinctive quantum optomechanical system \cite{Shahmoon2020} other than cavity QED setups \cite{Aspelmeyer2014}. In devising quantum heat engines using neutral atoms \cite{Brantut2013, Carollo2020, Myers2022, Koch2023, Nautiyal2024, Estrada2024}, a trapped-atom Otto engine can be benefited from PMDDIs to enhance the engine efficiency in finite-time many-body platforms \cite{Feyisa2024}, offering a renewed opportunity to explore the full potential of PMDDIs. Finally, a composite waveguide of two parallel optical nanofibers can reveal enhanced coupling efficiencies beyond the regimes using a single fiber waveguide \cite{Kien2022, Kien2024}, showing strong PMDDIs and entanglement generations, which can be utilized and designed for novel quantum interfaces. 

For the last remark on potential applications in quantum engineering of many-body entangled states, we note of possible controlled and tunable chiral coupling employed in an atom-nanophotonic cavity \cite{Hiew2023}, where strong light-atom couplings \cite{Samutpraphoot2020} can lead to atom-atom correlations \cite{Dordevic2021} and unexplored parameter regimes can arise for new applications in quantum technology. For example, scalable graph states generations via state carving technique \cite{Sorensen2003, Chen2015, Welte2017} has been proposed \cite{Chien2024}, where conditional detection events from reflected photons in an atom-nanophotonic cavity project the multiatom system into multiply entangled graph states \cite{Briegel2001, Hein2004}. These stationary qubits \cite{Cooper2024, Lanyon2013, Gong2019, Cao2023} are fundamental resource for one-way quantum computation \cite{Raussendorf2001, Raussendorf2003,Walther2005, Briegel2009}, which are also recently illustrated in reconfigurable atomic arrays \cite{Bluvstein2022, Bluvstein2024}, providing an alternative architecture of measurement-based quantum computing to their photonic counterparts \cite{Kiesel2005, Lu2007, Tokunaga2008, Schwartz2016, Larsen2019, Asavanant2019, Larsen2021, Thomas2022, Yang2022}. Ultimately, atomic arrays in optical tweezers \cite{Kaufman2012, Brown2019} becomes essential to tailor and control PMDDIs in free space or via coupling to waveguides, which opens a path to new possibilities of fundamental studies of collective light scattering and a new avenue for programmable quantum simulations \cite{Altman2021}. 

\section*{ACKNOWLEDGMENTS}

We acknowledge support from the National Science and Technology Council (NSTC), Taiwan, under the Grants No. 112-2112-M-001-079-MY3 and No. NSTC-112-2119-M-001-007, and from Academia Sinica under Grant AS-CDA-113-M04. We are also grateful for helpful discussions with T.-S. Gou, Y.-C. Wang, J.-S. You, and Y.-C. Chen, and support from TG 1.2 of NCTS at NTU.



\begin{thebibliography}{399}
\bibitem{Scully1997} M. O. Scully and M. S. Zubairy, {\it Quantum Optics} (Cambridge University Press, 1997).
\bibitem{Tannoudji1998} C. Cohen-Tannoudji, J. Dupont-Roc, and G. Grynberg, {\it Atom—Photon Interactions: Basic Process and Applications} (John Wiley \& Sons, Inc., 1998).
\bibitem{Loudon2000} R. Loudon, {\it The Quantum Theory of Light} (Oxford Science Publications, 2000). 
\bibitem{Breuer2002} H.-P. Breuer and F. Petruccione, {\it The Theory of Open Quantum Systems} (Oxford University Press, 2002). 
\bibitem{Walls2008} D. F. Walls and G. J. Milburn, {\it Quantum Optics} (Springer-Verlag Berlin, 2008). 
\bibitem{Tannoudji2011} C. Cohen-Tannoudji and D. Gu\'{e}ry-Odelin, {\it Advances in Atomic Physics: An overview} (World Scientific Publishing, 2011).
\bibitem{Bloch2008} I. Bloch, J. Dalibard, and W. Zwerger, Many-body physics with ultracold gases, Rev. Mod. Phys. {\bf 80}, 885 (2008). 
\bibitem{Giorgini2008} S. Giorgini, L. P. Pitaevskii, and S. Stringari, Theory of ultracold atomic Fermi gases, Rev. Mod. Phys. {\bf 80}, 1215 (2008). 
\bibitem{Metcalf1999} H. J. Metcalf and P. Straten, {\it Laser Cooling and Trapping} (Springer, 1999). 
\bibitem{McKay2011} D. C. McKay and B. DeMarco, Cooling in strongly correlated optical lattices: prospects and challenges, Rep. Prog. Phys. {\bf 74}, 054401 (2011). 
\bibitem{Eckardt2017} A. Eckardt, Colloquium: Atomic quantum gases in periodically driven optical lattices, Rev. Mod. Phys. {\bf 89}, 011004 (2017). 
\bibitem{Hammerer2010} K. Hammerer, A. S. Sørensen, and E. S. Polzik, Quantum interface between light and atomic ensembles, Rev. Mod. Phys. {\bf 82}, 1041 (2010). 
\bibitem{Lodahl2015} P. Lodahl, S. Mahmoodian, and S. Stobbe, Interfacing single photons and single quantum dots with photonic nanostructures, Rev. Mod. Phys. {\bf 87}, 347 (2015). 
\bibitem{Cirac1995} J. I. Cirac and P. Zoller, Quantum Computations with Cold Trapped Ions, Phys. Rev. Lett. {\bf 74}, 4091 (1995).
\bibitem{QIF} M. A. Nielsen and I. L. Chuang, {\it Quantum Computation and Quantum Information} (Cambridge University Press, 2000).
\bibitem{Buluta2009} I. Buluta and F. Nori, Quantum simulators, Science {\bf 326}, 108 (2009).
\bibitem{Georgescu2014} I. M. Georgescu, S. Ashhab, and F. Nori, Quantum simulation, Rev. Mod. Phys. {\bf 86}, 153 (2014). 
\bibitem{Altman2021} E. Altman {\it et al.}, Quantum Simulators: Architectures and Opportunities, PRX Quantum {\bf 2}, 017003 (2021). 
\bibitem{Stephen1964} M. J. Stephen, First-Order Dispersion Forces, J. Chem. Phys. {\bf 40}, 669 (1964).
\bibitem{Lehmberg1970} R. H. Lehmberg, Radiation from an $N$-Atom System. I. General Formalism, Phys. Rev. A {\bf 2}, 883 (1970). 
\bibitem{Jackson1998} D. J. Jackson, {\it Classical Electrodynamics} (John Wiley and Sons, New York, 1998). 
\bibitem{Zangwill2012} A. Zangwill, {\it Modern Electrodynamics} (Cambridge University Press, 2012). 
\bibitem{Dicke1954} R. H. Dicke, Coherence in spontaneous emission processes, Phys. Rev. {\bf 93}, 99 (1954). 
\bibitem{Gross1982} M. Gross and S. Haroche, Superradiance: An essay on the theory of collective spontaneous emission, Phys. Rep. {\bf 93}, 301 (1982). 
\bibitem{Devoe1996} R. G. DeVoe and R. G. Brewer, Observation of superradiant and subradiant spontaneous emission of two trapped ions. Phys. Rev. Lett. {\bf 76}, 2049 (1996).
\bibitem{Friedberg1973} R. Friedberg, S. R. Hartmann, and J. T. Manassah, Frequency shifts in emission and absorption by resonant systems ot two-level atoms, Phys. Rep. {\bf 7}, 101 (1973).
\bibitem{Scully2009} M. O. Scully, Collective lamb shift in single photon Dicke superradiance. Phys. Rev. Lett. 102, 143601 (2009).
\bibitem{Pellegrino2014} J. Pellegrino, R. Bourgain, S. Jennewein, Y. R. P. Sortais, A. Browaeys, S. D. Jenkins, and J. Ruostekoski, Observation of Suppression of Light Scattering Induced by Dipole-Dipole Interactions in a Cold-Atom Ensemble, Phys. Rev. Lett. {\bf 113}, 133602 (2014). 
\bibitem{Jennewein2016} S. Jennewein, M. Besbes, N. J. Schilder, S. D. Jenkins, C. Sauvan, J. Ruostekoski, J.-J. Greffet, Y. R. P. Sortais, and A. Browaeys, Coherent Scattering of Near-Resonant Light by a Dense Microscopic Cold Atomic Cloud, Phys. Rev. Lett. {\bf 116}, 233601 (2016). 
\bibitem{Roof2016} S. J. Roof, K. J. Kemp, M. D. Havey, and I. M. Sokolov, Observation of Single-Photon Superradiance and the Cooperative Lamb Shift in an Extended Sample of Cold Atoms, Phys. Rev. Lett. {\bf 117}, 073003 (2016). 
\bibitem{Jennewein2018} S. Jennewein, L. Brossard, Y. R. P. Sortais, A. Browaeys, P. Cheinet, J. Robert, and P. Pillet, Coherent scattering of near-resonant light by a dense, microscopic cloud of cold two-level atoms: Experiment versus theory, Phys. Rev. A {\bf 97}, 053816 (2018).
\bibitem{Guerin2016} W. Guerin, M. O. Araújo, and R. Kaiser, Subradiance in a Large Cloud of Cold Atoms, Phys. Rev. Lett. {\bf 116}, 083601 (2016).
\bibitem{Chomaz2012} L. Chomaz, L. Corman, T. Yefsah, R. Desbuquois, and J. Dalibard, Absorption imaging of a quasi-two-dimensional gas: a multiple scattering analysis, New J. Phys. {\bf 14}, 055001 (2012).
\bibitem{Andreoli2021} F. Andreoli, M. J. Gullans, A. A. High, A. Browaeys, and D. E. Chang, Maximum Refractive Index of an Atomic Medium, Phys. Rev. X {\bf 11}, 011026 (2021). 
\bibitem{Robicheaux2020} F. Robicheaux and R. T. Sutherland, Photon scattering from a cold, Gaussian atom cloud, Phys. Rev. A {\bf 101}, 013805 (2020).
\bibitem{Lin2022} C. Y. Lin and H. H. Jen, Interpretable machine-learning identification of the crossover from subradiance to superradiance in an atomic array, J. Phys. B: At. Mol. Opt. Phys. {\bf 55}, 135501 (2022).  
\bibitem{Rehler1971} N. E. Rehler and J. H. Eberly, Superradiance, Phys. Rev. A {\bf 3}, 1735 (1971). 
\bibitem{Shahmoon2017} E. Shahmoon, D. S. Wild, M. D. Lukin, and S. F. Yelin, Cooperative Resonances in Light Scattering from Two-Dimensional Atomic Arrays, Phys. Rev. Lett. {\bf 118}, 113601 (2017). 
\bibitem{Rui2020} J. Rui, D. Wei, A. Rubio-Abadal, S. Hollerith, J. Zeiher, D. M. Stamper-Kurn, C. Gross, and I. Bloch, A subradiant optical mirror formed by a single structured atomic layer, Nature {\bf 583}, 369 (2020).
\bibitem{Chang2012} D. E. Chang, L. Jiang, A. V. Gorshkov, and H. J. Kimble, Cavity QED with atomic mirrors, New J. of Phys. {\bf 14}, 063003 (2012).
\bibitem{Douglas2015} J. S. Douglas, H. Habibian, C.-L. Hung, A. V. Gorshkov, H. J. Kimble, and D. E. Chang, Quantum many-body models with cold atoms coupled to photonic crystals, Nat. Photon. {\bf 9}, 326 (2015).
\bibitem{Solano2017} P. Solano, P. Barberis-Blostein, F. K. Fatemi, L. A. Orozco, and S. L. Rolston, Super-radiance reveals infinite-range dipole interactions through a nanofiber Nat. commun. {\bf 8}, 1857 (2017).
\bibitem{Chang2018} D. E. Chang, J. S. Douglas, A. Gonz\'alez-Tudela, C.-L. Hung, H. J. Kimble, Colloquium: Quantum matter built from nanoscopic lattices of atoms and photons, Rev. Mod. Phys. {\bf 90}, 031002 (2018).
\bibitem{Mahmoodian2018} S. Mahmoodian, M. \ifmmode \check{C}\else \v{C}\fi{}epulkovskis, S. Das, P. Lodahl, K. Hammerer, and A. S. S\o{}rensen, Strongly Correlated Photon Transport in Waveguide Quantum Electrodynamics with Weakly Coupled Emitters, Phys. Rev. Lett. {\bf 121}, 143601 (2018).
\bibitem{Mahmoodian2020} S. Mahmoodian, G. Calajó, D. E. Chang, K. Hammerer, and A. S. Sørensen, Dynamics of Many-Body Photon Bound States in Chiral Waveguide QED, Phys. Rev. X {\bf 10}, 031011 (2020).
\bibitem{Kim2021} E. Kim, X. Zhang, V. S. Ferreira, J. Banker, J. K. Iverson, A. Sipahigil, M. Bello, A. Gonz\'alez-Tudela, M. Mirhosseini, and O. Painter, Quantum Electrodynamics in a Topological Waveguide, Phys. Rev. X {\bf 11}, 011015 (2021).
\bibitem{Fayard2021} N. Fayard, L. Henriet, A. Asenjo-Garcia, and D. E. Chang, Many-body localization in waveguide quantum electrodynamics, Phys. Rev. Res. {\bf 3}, 033233 (2021).
\bibitem{Sheremet2023} A. S. Sheremet, M. I. Petrov, I. V. Iorsh, A. V. Poshakinskiy, A. N. Poddubny, Waveguide quantum electrodynamics: collective radiance and photon-photon correlations, Rev. Mod. Phys. {\bf 95}, 015002 (2023). 
\bibitem{Tudela2024} A. González-Tudela, A. Reiserer, J. José García-Ripoll, and F. J. García-Vidal, Light–matter interactions in quantum nanophotonic devices, Nat. Rev. Phys. {\bf 6}, 166 (2024).
\bibitem{Perczel2017} J. Perczel, J. Borregaard, D. E. Chang, H. Pichler, S. F. Yelin, P. Zoller, and M. D. Lukin, Topological Quantum Optics in Two-Dimensional Atomic Arrays, Phys. Rev. Lett. {\bf 119}, 023603 (2017). 
\bibitem{Perczel2017_2} J. Perczel, J. Borregaard, D. E. Chang, H. Pichler, S. F. Yelin, P. Zoller, and M. D. Lukin, Photonic band structure of two-dimensional atomic lattices, Phys. Rev. A {\bf 96}, 063801 (2017). 
\bibitem{Perczel2020} J. Perczel, J. Borregaard, D. E. Chang, S. F. Yelin, and M. D. Lukin, Topological Quantum Optics Using Atomlike Emitter Arrays Coupled to Photonic Crystals, Phys. Rev. Lett. {\bf 124}, 083603 (2020).
\bibitem{Meiser2009} D. Meiser, Jun Ye, D. R. Carlson, and M. J. Holland, Prospects for a Millihertz-Linewidth Laser, Phys. Rev. Lett. {\bf 102}, 163601 (2009). 
\bibitem{Bohnet2012} J. G. Bohnet, Z. Chen, J. M. Weiner, D. Meiser, M. J. Holland, and J. K. Thompson, A steady-state superradiant laser with less than one intracavity photon, Nature {\bf 484}, 78 (2012).
\bibitem{Garcia2017} A. Asenjo-Garcia, M. Moreno-Cardoner, A. Albrecht, H. J. Kimble, and D. E. Chang, Exponential improvement in photon storage fidelities using subradiance and “Selective Radiance” in atomic arrays, Phys. Rev. X {\bf 7}, 031024 (2017).
\bibitem{Bekenstein2020} R. Bekenstein, I. Pikovski, H. Pichler, E. Shahmoon, S. F. Yelin, and M. D. Lukin, Quantum metasurfaces with atom arrays, Nat. Phys. 16, 676–681 (2020).
\bibitem{Hutson2024} R. B. Hutson, W. R. Milner, L. Yang, J. Ye, and C. Sanner, Observation of millihertz-level cooperative Lamb shifts in an optical atomic clock, Science {\bf 383}, 384 (2024). 
\bibitem{Holzinger2024} R. Holzinger, J. S. Peter, S. Ostermann, H. Ritsch, and S. Yelin, Harnessing quantum emitter rings for efficient energy transport and trapping, Optica Quantum {\bf 2}, 57 (2024). 
\bibitem{Lukin2003} M. D. Lukin, Colloquium: Trapping and manipulating photon states in atomic ensembles, Rev. Mod. Phys. {\bf 75}, 457 (2003). 
\bibitem{Fleischhauer2005} M. Fleischhauer, A. Imamoglu, and J. P. Marangos, Electromagnetically induced transparency: Optics in coherent media, Rev. Mod. Phys. {\bf 77}, 633 (2005).
\bibitem{Breuer2012} H.-P. Breuer, Foundations and measures of quantum non-Markovianity, J. Phys. B: At. Mol. Opt. Phys. {\bf 45}, 154001 (2012). 
\bibitem{Rivas2014} Á. Rivas, S. F. Huelga, and M. B Plenio, Quantum non-Markovianity: characterization, quantification and detection, Rep. Prog. Phys. {\bf 77}, 094001 (2014). 
\bibitem{Breuer2016} H.-P. Breuer, E.-M. Laine, J. Piilo, and B. Vacchini, Colloquium: Non-Markovian dynamics in open quantum systems, Rev. Mod. Phys. {\bf 88}, 021002 (2016). 
\bibitem{Defenu2023} N. Defenu, T. Donner, T. Macrì, G. Pagano, S. Ruffo, and A. Trombettoni, Long-range interacting quantum systems, Rev. Mod. Phys. {\bf 95}, 035002 (2023). 
\bibitem{Leibfried2003} D. Leibfried, R. Blatt, C. Monroe, and D. Wineland, Quantum dynamics of single trapped ions, Rev. Mod. Phys. {\bf 75}, 281 (2003). 
\bibitem{Monroe2021} C. Monroe, W. C. Campbell, L.-M. Duan, Z.-X. Gong, A. V. Gorshkov, P. W. Hess, R. Islam, K. Kim, N. M. Linke, G. Pagano, P. Richerme, C. Senko, and N. Y. Yao, Programmable quantum simulations of spin systems with trapped ions, Rev. Mod. Phys. {\bf 93}, 025001 (2021).
\bibitem{Periwal2021} A. Periwal, E. S. Cooper, P. Kunkel, J. F. Wienand, E. J. Davis, and M. Schleier-Smith, Programmable interactions and emergent geometry in an array of atom clouds, Nature {\bf 600}, 630 (2021). 
\bibitem{Saffman2010} M. Saffman, T. G. Walker, and K. Mølmer, Quantum information with Rydberg atoms, Rev. Mod. Phys. {\bf 82}, 2313 (2010).
\bibitem{Hung2016} C.-L. Hung, A. Gonz\'alez-Tudela, J. I. Cirac, and H. J. Kimble, Quantum spin dynamics with pairwise-tunable, long-range interactions, Proc. Natl. Acad. Sci. {\bf 113}, E4946 (2016).
\bibitem{Karzig2015} T. Karzig, C.-E. Bardyn, N. H. Lindner, and G. Refael, Topological Polaritons, Phys. Rev. X {\bf 5}, 031001 (2015). 
\bibitem{Ozawa2019} T. Ozawa, H. M. Price, A. Amo, N. Goldman, M. Hafezi, L. Lu, M. C. Rechtsman, D. Schuster, J. Simon, O. Zilberberg, and I. Carusotto, Topological photonics, Rev. Mod. Phys. {\bf 91}, 015006 (2019). 
\bibitem{Ma2019} G. Ma, M. Xiao, and C. T. Chan, Topological phases in acoustic and mechanical systems, Nat. Rev. Phys. {\bf 1}, 281 (2019).
\bibitem{Cooper2019} N. R. Cooper, J. Dalibard, and I. B. Spielman, Topological bands for ultracold atoms, Rev. Mod. Phys. {\bf 91}, 015005 (2019). 
\bibitem{Weisskopf1930} V. Weisskopf and E. Wigner, Z. Phys. {\bf 63}, 54 (1930). 
\bibitem{Steck_alkali} D. A. Steck, Alkali D Line Data, https://steck.us/alkalidata/
\bibitem{Jen2019_2D} H. H. Jen, Super- and sub-radiance from two-dimensional resonant dipole-dipole interactions, Sci. Rep. {\bf 9}, 5804 (2019). 
\bibitem{Jen2020_book} H. H. Jen, {\it Collective Light Emission: Many quantum emitters} (IOP Publishing, 2020). 
\bibitem{Yaghjian1980} A. D. Yaghjian, Electric Dyadic Green‘s Functions in the Source Region, Proceedings of the IEEE {\bf 68}, 248 (1980).
\bibitem{Maximo2015} C. E. M´aximo, N. Piovella, Ph. W. Courteille, R. Kaiser, and R. Bachelard, Spatial and temporal localization of light in two dimensions, Phys. Rev. A {\bf 92}, 062702 (2015). 
\bibitem{Longo2016} P. Longo, C. H. Keitel, and J. Evers, Tailoring superradiance to design artificial quantum systems, Sci. Rep. {\bf 6}, 23628 (2016).
\bibitem{Tudela2015} A. Gonz\'{a}lez-Tudela, C.-L. Hung, D. E. Chang, J. I. Cirac, and H. J. Kimble, Subwavelength vacuum lattices and atom–atom interactions in two-dimensional photonic crystals, Nat. Photonics {\bf 9}, 320 (2015).
\bibitem{Tecer2024} M. Te\ifmmode \check{c}\else \v{c}\fi{}er, M. Di Liberto, P. Silvi, S. Montangero, F. Romanato, and G. Calaj\'o, Strongly Interacting Photons in 2D Waveguide QED, Phys. Rev. Lett. {\bf 132}, 163602 (2024). 
\bibitem{Tudela2013} A. Gonz\'alez-Tudela and D. Porras, Mesoscopic Entanglement Induced by Spontaneous Emission in Solid-State Quantum Optics, Phys. Rev. Lett. {\bf 110}, 080502 (2013).
\bibitem{Lodahl2017} P. Lodahl, S. Mahmoodian, S. Stobbe, A. Rauschenbeutel, P. Schneeweiss, J. Volz, H. Pichler, and P. Zoller, Chiral quantum optics, Nature {\bf 541}, 473 (2017).
\bibitem{Arauju2017} M. o. Araújo, W. Guerin, and R. Kaiser, Decay dynamics in the coupled-dipole model. J. Modern Opt. {\bf 65}, 1345 (2017).
\bibitem{Jenkins2012} S. D. Jenkins and J. Ruostekoski, Controlled manipulation of light by cooperative response of atoms in an optical lattice, Phys. Rev. A {\bf 86}, 031602(R) (2012). 
\bibitem{Jenkins2016} S. D. Jenkins, J. Ruostekoski, J. Javanainen, S. Jennewein,R. Bourgain, J. Pellegrino, Y. R. P. Sortais, and A. Browaeys, Collective resonance fluorescence in small and dense atom clouds: Comparison between theory and experiment, Phys.Rev. A 94, 023842 (2016).
\bibitem{Sutherland2016_1} R. T. Sutherland and F. Robicheaux, Coherent forward broadening in cold atom clouds, Phys. Rev. A {\bf 93}, 023407 (2016). 
\bibitem{Sutherland2016_2} R. T. Sutherland and F. Robicheaux, Collective dipole-dipole interactions in an atomic array, Phys. Rev. A {\bf 94}, 013847 (2016).
\bibitem{Zhu2016} B. Zhu, J. Cooper, J. Ye, and A. M. Rey, Light scattering from dense cold atomic media, Phys. Rev. A {\bf 94}, 023612 (2016).
\bibitem{Ferioli2021} G. Ferioli, A. Glicenstein, F. Robicheaux, R. T. Sutherland, A. Browaeys, and I. Ferrier-Barbut, Laser-Driven Superradiant Ensembles of Two-Level Atoms near Dicke Regime, Phys. Rev. Lett. {\bf 127}, 243602 (2021). 
\bibitem{Bromley2016} S. L. Bromley, B. Zhu, M. Bishof, X. Zhang, T. Bothwell, J.
Schachenmayer, T. L. Nicholson, R. Kaiser, S. F. Yelin, M. D. Lukin, A. M. Rey, and J. Ye, Collective atomic scattering and motional effects in a dense coherent medium, Nat. Commun. {\bf 7}, 11039 (2016).
\bibitem{Jen2018_helical} H. H. Jen, M. S. Chang, and Y. C. Chen, Cooperative light scattering from helical-phase-imprinted atomic rings, Sci. Rep. {\bf 8}, 9570 (2018). 
\bibitem{Bettles2016} R. J. Bettles, S. A. Gardiner, and C. S. Adams, Enhanced Optical Cross Section via Collective Coupling of Atomic Dipoles in a 2D Array, Phys. Rev. Lett. {\bf 116}, 103602 (2016).
\bibitem{Lei2023} M. Lei, R. Fukumori, J. Rochman, B. Zhu, M. Endres, J. Choi, and A. Faraon, Many-body cavity quantum electrodynamics with driven inhomogeneous emitters, Nature {\bf 617}, 271 (2023). 
\bibitem{Scully2015} M. O. Scully, Single Photon Subradiance: Quantum Control of Spontaneous Emission and Ultrafast Readout, Phys. Rev. Lett. {\bf 115}, 243602 (2015).
\bibitem{Facchinetti2016} G. Facchinetti, S. D. Jenkins, and J. Ruostekoski, Storing Light with Subradiant Correlations in Arrays of Atoms, Phys. Rev. Lett. {\bf 117}, 243601 (2016).
\bibitem{Plankensteiner2015} D. Plankensteiner, L. Ostermann, H. Ritsch, and C. Genes, Selective protected state preparation of coupled dissipative quantum emitters, Sci. Rep. {\bf 5}, 16231 (2015).
\bibitem{Jen2016_SR} H. H. Jen, M.-S. Chang, and Y.-C. Chen, Cooperative single-photon subradiant states, Phys. Rev. A {\bf 94}, 013803 (2016).
\bibitem{Jen2016_SR2} H. H. Jen, Cooperative single-photon subradiant states in a three-dimensional atomic array, Ann. Phys. (N. Y.) {\bf 374}, 27 (2016).
\bibitem{Hsiao2018} Y.-F. Hsiao, P.-J. Tsai, H.-S. Chen, S.-X. Lin, C.-C. Hung, C.-H. Lee, Y.-H. Chen, Y.-F. Chen, I. A. Yu, and Y.-C. Chen, Highly Efficient Coherent Optical Memory Based on Electromagnetically Induced Transparency, Phys. Rev. Lett. {\bf 120}, 183602 (2018). 
\bibitem{Plankensteiner2017} D. Plankensteiner, C. Sommer, H. Ritsch, and C. Genes, Cavity Antiresonance Spectroscopy of Dipole Coupled Subradiant Arrays, Phys. Rev. Lett. {\bf 119}, 093601 (2017).
\bibitem{Jenkins2017} S. D. Jenkins, J. Ruostekoski,N. Papasimakis, S. Savo,and N. I. Zheludev, Many-Body Subradiant Excitations in Metamaterial Arrays: Experiment and Theory, Phys. Rev. Lett. {\bf 119}, 053901 (2017).
\bibitem{Albrecht2019} A. Albrecht, L. Henriet, A. Asenjo-Garcia, P. B Dieterle, O. Painter, and D. E. Chang, Subradiant states of quantum bits coupled to a one-dimensional waveguide, New J. Phys. {\bf 21}, 025003 (2019).
\bibitem{Zhang2019} Y.-X. Zhang and K. Mølmer, Theory of Subradiant States of a One-Dimensional Two-Level Atom Chain, Phys. Rev. Lett. {\bf 122}, 203605 (2019). 
\bibitem{Jen2017_MP} H. H. Jen, Phase-imprinted multiphoton subradiant states, Phys. Rev. A {\bf 96}, 023814 (2017).
\bibitem{Jen2018_directional} H. H. Jen, Directional subradiance from helical-phase-imprinted multiphoton states, Sci. Rep. {\bf 8}, 7163 (2018). 
\bibitem{Williamson2020} L. A. Williamson and J. Ruostekoski, Optical response of atom chains beyond the limit of low light intensity: The validity of the linear classical oscillator model, Phys. Rev. Res. {\bf 2}, 023273(2020).
\bibitem{Robicheaux2021} F. Robicheaux and D. A. Suresh, Beyond lowest order mean-field theory for light interacting with atom arrays, Phys. Rev. A {\bf 104}, 023702 (2021).
\bibitem{Robicheaux2023} F. Robicheaux and D. A. Suresh, Intensity effects of light coupling to one- or two-atom arrays of infinite extent, Phys. Rev. A {\bf 108}, 013711 (2023).
\bibitem{Bigorda2023} O. Rubies-Bigorda, S. Ostermann, and S. F. Yelin, Characterizing superradiant dynamics in atomic arrays via a cumulant expansion approach, Phys. Rev. Res. {\bf 5}, 013091 (2023).
\bibitem{Kusmierek2023} K. J. Kusmierek, S. Mahmoodian, M. Cordier, J. Hinney, A. Rauschenbeutel, M. Schemmer, P. Schneeweiss, J. Volz, and K. Hammerer, Higher-order mean-field theory of chiral waveguide QED, SciPost Phys. Core {\bf 6}, 041 (2023).
\bibitem{Ferioli2024} G. Ferioli, S. Pancaldi, A. Glicenstein, D. Cl\'{e}ment, A. Browaeys, and I. Ferrier-Barbut, Non-Gaussian Correlations in the Steady State of Driven-Dissipative Cloudsof Two-Level Atoms, Phys. Rev. Lett. {\bf 132}, 133601 (2024). 
\bibitem{Funo2024} K. Funo and A. Ishizaki, Dynamics of a Quantum System Interacting withWhite Non-Gaussian Baths: Poisson Noise Master Equation, Phys. Rev. Lett. {\bf 132}, 170402 (2024). 
\bibitem{Wang2024} C.-H. Wang, N.-Y. Tsai, Y.-C. Wang, and H. H. Jen, Light scattering properties beyond weak-field excitation in atomic ensembles, Phys. Rev. A {\bf 110}, 013708 (2024). 
\bibitem{Corman2017} L. Corman, J. L. Ville, R. Saint-Jalm, M. Aidelsburger, T. Bienaim\'{e}, S. Nascimb`ene, J. Dalibard, and J. Beugnon, Transmission of near-resonant light through a dense slab of cold atoms, Phys. Rev. A {\bf 96}, 053629 (2017).
\bibitem{Javanainen2019} J. Javanainen and R. Rajapakse, Light propagation in systems involving two-dimensional atomic lattices, Phys. Rev. A {\bf 100}, 013616 (2019).
\bibitem{Ballantine2021} K. E. Ballantine and J. Ruostekoski, Quantum Single-Photon Control, Storage, and Entanglement Generation with Planar Atomic Arrays, PRX Quantum {\bf 2}, 040362 (2021).
\bibitem{Shah2024} F. Shah, T. L. Patti, O. Rubies-Bigorda, and S. F. Yelin, Quantum computing with subwavelength atomic arrays, Phys. Rev. A {\bf 109}, 012613 (2024).
\bibitem{Robicheaux2024} F. Robicheaux, Spatial averaging for light reflection and transmission through cold atom arrays, arXiv:2410.18855 (2024). 
\bibitem{Meystre2007} P. Meystre and M. Sargent, {\it Elements of quantum optics} (Springer, 2007). 
\bibitem{Caneva2015} T. Caneva, M. T. Manzoni, T. Shi, J. S. Douglas, J. I. Cirac, and D. E. Chang, Quantum dynamics of propagating photons with strong interactions: a generalized input–output formalism, New J. Phys. {\bf 17}, 113001 (2015). 
\bibitem{Fan2010} S. Fan, and S. E. Kocaba\ifmmode \mbox{\c{s}}\else \c{s}\fi{}, and J.-T. Shen, Input-output formalism for few-photon transport in one-dimensional nanophotonic waveguides coupled to a qubit, Phys. Rev. A {\bf 82}, 063821 (2010).
\bibitem{Shi2011} T. Shi, S. Fan, and C. P. Sun, Two-photon transport in a waveguide coupled to a cavity in a two-level system, Phys. Rev. A {\bf 84}, 063803 (2011). 
\bibitem{Pletyukhov2012} M. Pletyukhov and V. Gritsev, Scattering of massless particles in one-dimensional chiral channel, New. J. Phys. {\bf 14}, 095028 (2012). 
\bibitem{Xu2015} S. Xu and S. Fan, Input-output formalism for few-photon transport: A systematic treatment beyond two photons, Phys. Rev. A {\bf 91}, 043845 (2015).
\bibitem{Jones2020} R. Jones, G. Buonaiuto, B. Lang, I. Lesanovsky, and B. Olmos, Collectively Enhanced Chiral Photon Emission from an Atomic Array near a Nanofiber, Phys. Rev. Lett. {\bf 124}, 093601 (2020).
\bibitem{Scully2006} M. O. Scully, E. S. Fry, C. H. Raymond Ooi, and K. Wódkiewicz, Directed Spontaneous Emission from an Extended Ensemble of $N$ Atoms: Timing Is Everything, Phys. Rev. Lett. {\bf 96}, 010501 (2006).
\bibitem{Mazets2007} I. E. Mazets and G. Kurizki, Multiatom cooperative emission following single-photon absorption: Dicke-state dynamics, J. Phys. B: At. Mol. Opt. Phys. {\bf 40}, F105 (2007). 
\bibitem{Jen2015} H. H. Jen, Superradiant cascade emissions in an atomic ensemble via four-wave mixing, Ann. of Phys. (N.Y.) {\bf 360}, 556 (2015).
\bibitem{Hsu2024} T. Hsu, K.-T. Lin, and G.-D. Lin, Cooperative states and shift in resonant scattering of an atomic ensemble, New Journal of Physics {\bf 26}, 053206 (2024). 
\bibitem{Rohlsberger2010} R. R\"{o}hlsberger, K. Schlage, B. Sahoo, S. Couet, and R. R\"{u}ffer, Collective Lamb shift in single-photon superradiance, Science {\bf 328}, 1248 (2010).
\bibitem{Keaveney2012} J. Keaveney, A. Sargsyan, U. Krohn, I. G. Hughes, D. Sarkisyan, and C. S. Adams, Cooperative Lamb shift in an atomic vapor layer of nanometer thickness, Phys. Rev. Lett. {\bf 108}, 173601 (2012).
\bibitem{Peyrot2018} T. Peyrot, Y. R. P. Sortais, A. Browaeys, A. Sargsyan, D. Sarkisyan, J. Keaveney, I. G. Hughes, and C. S. Adams, Collective Lamb shift of a nanoscale atomic vapor layer within a sapphire cavity, Phys. Rev. Lett. {\bf 120}, 243401 (2018). 
\bibitem{Meir2014} Z. Meir, O. Schwartz, E. Shahmoon, D. Oron, and R. Ozeri, Cooperative Lamb shift in a mesoscopic atomic array, Phys. Rev. Lett. {\bf 113}, 193002 (2014).
\bibitem{Wen2019} P. Y. Wen, K.-T. Lin, A. F. Kockum, B. Suri, H. Ian, J.C. Chen, S. Y. Mao, C. C. Chiu, P. Delsing, F. Nori, G.-D. Lin, and I.-C. Hoi, Large Collective Lamb Shift of Two Distant Superconducting Artificial Atoms, Phys. Rev. Lett. {\bf 123}, 233602 (2019). 
\bibitem{Hebenstreit2017} M. Hebenstreit, B. Kraus, L. Ostermann, and H. Ritsch, Subradiance via entanglement in atoms with several independent decay channels. Phys. Rev. Lett. {\bf 118}, 143602 (2017).
\bibitem{Parmee2020} C. D. Parmee and J. Ruostekoski, Signatures of optical phase transitions in superradiant and subradiant atomic arrays, Commun. Phys. {\bf 3}, 205 (2020). 
\bibitem{Jen2018_SR1} H. H. Jen, M.-S. Chang, and Y.-C. Chen, Cooperative light scattering from helical-phase-imprinted atomic rings, Sci. Rep. {\bf 8}, 9570 (2018).
\bibitem{Jen2018_SR2} H. H. Jen, Directional subradiance from helical-phase-imprinted multiphoton states, Sci. Rep. {\bf 8}, 7163 (2018).
\bibitem{Moreno-Cardoner2019} M. Moreno-Cardoner, D. Plankensteiner, L. Ostermann, D. E. Chang, and H. Ritsch, Subradiance-enhanced excitation transfer between dipole-coupled nanorings of quantum emitters, Phys. Rev. A {\bf 100}, 023806 (2019).
\bibitem{Moreno-Cardoner2022} M. Moreno-Cardoner, R. Holzinger, and H. Ritsch, Efficient nano-photonic antennas based on dark states in quantum emitter rings, Opt. Exp. {\bf 30}, 10779 (2022). 
\bibitem{Facchinetti_2016} G. Facchinetti, S. D. Jenkins, and J. Ruostekoski, Storing light with subradiant correlations in arrays of atoms, Phys. Rev. Lett. {\bf 117}, 243601 (2016). 
\bibitem{Bhatti2018} D. Bhatti, R. Schneider, S. Oppel, and J. von Zanthier, Directional Dicke Subradiance with Nonclassical and Classical Light Sources, Phys. Rev. Lett. {\bf 120}, 113603 (2018).
\bibitem{Ferioli2021_sub} G. Ferioli, A. Glicenstein, L. Henriet, I. Ferrier-Barbut, and A. Browaeys, Storage and Release of Subradiant Excitations in a Dense Atomic Cloud, Phys. Rev. X {\bf 11}, 021031 (2021). 
\bibitem{Guimond2019} P.-O. Guimond, A. Grankin, D. V. Vasilyev, B. Vermersch, and P. Zoller, Subradiant Bell states in distant atomic arrays, Phys. Rev. Lett. {\bf 122}, 093601 (2019). 
\bibitem{Pineiro2019} A. Pi\~neiro Orioli and A. M. Rey, Dark States of Multilevel Fermionic Atoms in Doubly Filled Optical Lattices, Phys. Rev. Lett. {\bf 123}, 223601 (2019). 
\bibitem{Needham2019} J. A. Needham, I. Lesanovsky, and B. Olmos, Subradiance-protected excitation transport, New J. Phys. {\bf 21}, 073061 (2019).
\bibitem{Kubo1962} R. Kubo, Generalized cumulant expansion method, J. Phys. Soc. Jpn. {\bf 17}, 1100 (1962).
\bibitem{Jen2012} H. H. Jen, Positive-$P$ phase-space-method simulation of superradiant emission from a cascade atomic ensemble, Phys. Rev. A {\bf 85}, 013835 (2012).
\bibitem{Jen2022_EIT} H. H. Jen, G.-D. Lin, and Y.-C. Chen, Resonant dipole-dipole interactions in electromagnetically induced transparency, Phys. Rev. A {\bf 105}, 063711 (2022). 
\bibitem{Sutherland2017} R. T. Sutherland and F. Robicheaux, Superradiance in inverted multilevel atomic clouds, Phys. Rev. A {\bf 95}, 033839 (2017).
\bibitem{Liedl2024} C. Liedl, F. Tebbenjohanns, C. Bach, S. Pucher, A. Rauschenbeutel, and P. Schneeweiss, Observation of Superradiant Bursts in a Cascaded Quantum System, Phys. Rev. X {\bf 14}, 011020 (2024). 
\bibitem{Bach2024} C. Bach, F. Tebbenjohanns, C. Liedl, P. Schneeweiss, and A. Rauschenbeutel, Emergence of second-order coherence in superfluorescence, arXiv: 2407.12549 (2024). 
\bibitem{Tebbenjohanns2024} F. Tebbenjohanns, C. Bach, A. Rauschenbeutel, C. D. Mink and M. Fleischhauer, Predicting correlations in superradiant emission from a cascaded quantum system, arXiv: 2407.02154 (2024). 
\bibitem{Masson2020} S. J. Masson, I. Ferrier-Barbut, L. A. Orozco, A. Browaeys, and A. Asenjo-Garcia, Many-Body Signatures of Collective Decay in Atomic Chains, Phys. Rev. Lett. {\bf 125}, 263601 (2020). 
\bibitem{Masson2022} S. J. Masson and A. Asenjo-Garcia, Universality of Dicke superradiance in arrays of quantum emitters, Nat. Commun. {\bf 13}, 2285 (2022). 
\bibitem{Masson2024} S. J. Masson, J. P. Covey, S. Will, and A. Asenjo-Garcia, Dicke Superradiance in Ordered Arrays of Multilevel Atoms, PRX Quantum {\bf 5}, 010344 (2024).
\bibitem{Bigorda2022} O. Rubies-Bigorda and S. F. Yelin, Superradiance and subradiance in inverted atomic arrays, Phys. Rev. A {\bf 106}, 053717 (2022). 
\bibitem{Ferioli2023} G. Ferioli, A. Glicenstien, I. Ferrir-Barbut, and A. Browaeys, A non-equilibrium superradiant phase transition in free space, Nat. Phys. {\bf 19}, 1345 (2023). 
\bibitem{Hazzard2014} K. R. Hazzard, B. Gadway, M. Foss-Feig, B. Yan, S. A. Moses, J. P. Covey, N. Y. Yao, M. D. Lukin, J. Ye, D. S. Jin, and A. M. Rey, Many-Body Dynamics of Dipolar Molecules in an Optical Lattice, Phys. Rev. Lett. {\bf 113}, 195302 (2014). 
\bibitem{Agarwal2024} S. Agarwal, E. Chaparro, D. Barberena, A. P. Orioli, G. Ferioli, S. Pancaldi, I. Ferrier-Barbut, A. Browaeys, and A. M. Rey, Directional superradiance in a driven ultracold atomic gas in free-space, arXiv:2403.15556 (2024). 
\bibitem{Mink2022} C. D. Mink, D. Petrosyan and M. Fleischhauer, Hybrid discrete-continuous truncated Wigner approximation for driven, dissipative spin systems, Phys. Rev. Res. {\bf 4}, 043136 (2022). 
\bibitem{Mink2023} C. D. Mink and M. Fleischhauer, Collective radiative interactions in the discrete truncated Wigner approximation, SciPost Phys. {\bf 15}, 233 (2023).
\bibitem{Gardiner2004} C. W. Gardiner, {\it Handbook of Stochastic Methods: for Physics, Chemistry and the Natural Sciences} (Springer-Verlag Berlin, 2004). 
\bibitem{Haken1970} H. Haken, {\it Laser Theory} (Springer-Verlag Berlin, 1970). 
\bibitem{Gardiner2000} C. W. Gardiner and P. Zoller, {\it Quantum Noise: A Handbook of Markovian and Non-Markovian Quantum Stochastic Methods with Applications to Quantum Optics}, 2nd ed. (Springer-Verlag Berlin, 2000). 
\bibitem{Drummond1980} P. D. Drummond and C. W. Gardiner, Generalised P-representations in quantum optics, J. Phys. A {\bf 13}, 2353 (1980). 
\bibitem{Drummond1987} P. D. Drummond and S. J. Carter, Quantum-field theory of squeezing in solitons, J. Opt. Soc. Am. B {\bf 4}, 1565 (1987).
\bibitem{Drummond1991} P. D. Drummond and M. G. Raymer, Quantum theory of propagation of nonclassical radiation in a near-resonant medium, Phys. Rev. A {\bf 44}, 2072 (1991). 
\bibitem{Kiesewetter2023} S. Kiesewetter, R. R. Joseph and P. D. Drummond, xSPDE3: Extensible software for stochastic ordinary and partial differential equations, SciPost Phys. Codebases {\bf 17} (2023). 
\bibitem{Horvath2024} D. X. Horváth and C. Rylands, Full counting statistics of charge in quenched quantum gases, Phys. Rev. A {\bf 109}, 043302 (2024).
\bibitem{Vogt1996} A. W. Vogt, J. I. Cirac, and P. Zoller, Collective laser cooling of two trapped ions, Phys. Rev. A {\bf 53}, 950 (1996).
\bibitem{Xu2016} M. Xu, S. B. Jäger, S. Schütz, J. Cooper, G. Morigi, and M. J. Holland, Supercooling of Atoms in an Optical Resonator, Phys. Rev. Lett. {\bf 116}, 153002 (2016).
\bibitem{Maximo2018} C. E. M\'{a}ximo, R. Bachelard, and R. Kaiser, Optical binding with cold atoms, Phys. Rev. A {\bf 97}, 043845 (2018).
\bibitem{Gisbert2019} A. T. Gisbert, N. Piovella, and R. Bachelard, Stochastic heating and self-induced cooling in optically bound pairs of atoms, Phys. Rev. A {\bf 99}, 013619 (2019).
\bibitem{Wang2023} C.-H. Wang, Y.-C. Wang, C.-C. Chen, C.-C. Wang, and H. H. Jen, Enhanced dark-state sideband cooling in trapped atoms via photon-mediated dipole-dipole interactions, Phys. Rev. A {\bf 107}, 023117 (2023).
\bibitem{Bigorda2024} O. Rubies-Bigorda, R. Holzinger, A. Asenjo-Garcia, O. Romero-Isart, H. Ritsch, S. Ostermann, C. Gonzalez-Ballestero, S. F. Yelin, C. C. Rusconi, Collectively enhanced ground-state cooling in subwavelength atomic arrays, arXiv:2405.18482 (2024). 
\bibitem{Haake1993} F. Haake, M. I. Kolobov, C. Fabre, E. Giacobino, and S. Reynaud, Superradiant laser, Phys. Rev. Lett. {\bf 71}, 995 (1993).
\bibitem{Meiser2010} D. Meiser and M. J. Holland, Steady-state superradiance with alkaline-earth-metal atoms, Phys. Rev. A {\bf 81}, 033847 (2010).
\bibitem{Maier2014} T. Maier, S. Kraemer, L. Ostermann, and H. Ritsch, A superradiant clock laser on a magic wavelength optical lattice, Opt. Express {\bf 22}, 13269 (2014).
\bibitem{Ludlow2015} A. D. Ludlow, M. M. Boyd, J. Ye, E. Peik, and P. O.Schmidt, Optical atomic clocks, Rev. Mod. Physics {\bf 87}, 637 (2015). 
\bibitem{Jen2016_SL} H. H. Jen, Superradiant laser: Effect of long-range dipole-dipole interaction, Phys. Rev. A {\bf 94}, 053813 (2016). 
\bibitem{Akkermans2008} E. Akkermans, A. Gero, and R. Kaiser, Photon Localization and Dicke Superradiance in Atomic Gases, Phys. Rev. Lett. {\bf 101}, 103602 (2008). 
\bibitem{Bienaime2012} Tom Bienaimé, Nicola Piovella, and Robin Kaiser, Controlled Dicke Subradiance from a Large Cloud of Two-Level Systems, Phys. Rev. Lett. {\bf 108}, 123602 (2012). 
\bibitem{Sollner2015} I. Söllner, S. Mahmoodian, S. L. Hansen, L. Midolo, G.Kirsanske, T. Pregnolato, H. El-Ella, E. H. Lee, J. D. Song,S. Stobbe, and P. Lodahl, Deterministic photon–emitter coupling in chiral photonic circuits, Nat. Nanotechnol. {\bf 10}, 775 (2015).
\bibitem{Young2015} A. B. Young, A. C. T. Thijssen, D. M. Beggs, P.Androvitsaneas, L. Kuipers, J. G. Rarity, S. Hughes, and R. Oulton, Polarization Engineering in Photonic Crystal Waveguides for Spin-Photon Entanglers, Phys. Rev. Lett. {\bf 115}, 153901 (2015).
\bibitem{Vetsch2010} E. Vetsch, D. Reitz, G. Sagué, R. Schmidt, S. T. Dawkins, and A. Rauschenbeutel, Optical Interface Created by Laser-Cooled Atoms Trapped in the Evanescent Field Surrounding an Optical Nanofiber, Phys. Rev. Lett. {\bf 104}, 203603 (2010). 
\bibitem{Corzo2019} N. V. Corzo, J. Raskop, A. Chandra, A. S. Sheremet, B. Gouraud, and J. Laurat, Waveguide-coupled single collective excitation of atomic arrays, Nature {\bf 566}, 359 (2019).
\bibitem{Johnson2019} A. Johnson, M. Blaha, A. E. Ulanov, A. Rauschenbeutel, P. Schneeweiss, and J. Volz, Observation of Collective Superstrong Coupling of Cold Atoms to a 30-m Long Optical Resonator, Phys. Rev. Lett. {\bf 123}, 243602 (2019).
\bibitem{Bliokh2014} K. Y. Bliokh, A. Y. Bekshaev, and F. Nori, Extraordinary momentum and spin in evanescent waves, Nat. Commun. {\bf 5}, 3300 (2014).
\bibitem{Bliokh2015} K. Y. Bliokh and F. Nori, Transverse and longitudinal angular momenta of light, Phys. Rep. {\bf 592}, 1 (2015).
\bibitem{Kien2005} F. Le Kien, S. D. Gupta, K. P. Nayak, and K. Hakuta, Nanofiber-mediated radiative transfer between two distant atoms. Phys. Rev. A {\bf 72}, 063815 (2005).
\bibitem{Kien2008} F. Le Kien and K. Hakuta, Cooperative enhancement of channeling of emission from atoms into a nanofiber. Phys. Rev. A {\bf 77}, 013801 (2008).
\bibitem{Loo2013} A. F. van Loo , A. Fedorov, K. Lalumière, B. C. Sanders, A. Blais, and A. Wallraff, Photon-Mediated Interactions Between Distant Artificial Atoms, Science {\bf 342}, 1494 (2013). 
\bibitem{Goban2015} A. Goban, C.-L. Hung, J. D. Hood, S.-P. Yu, J. A. Muniz, O. Painter, and H. J. Kimble, Superradiance for Atoms Trapped along a Photonic Crystal Waveguide, Phys. Rev. Lett. {\bf 115}, 063601 (2015).  
\bibitem{Pichler2015} H. Pichler, T. Ramos, A. J. Daley, and P. Zoller, Quantum optics of chiral spin networks, Phys. Rev. A {\bf 91}, 042116 (2015).
\bibitem{Shahmoon2016} E. Shahmoon, P. Grišins, H. P. Stimming, I. Mazets, and G. Kurizki, Highly nonlocal optical nonlinearities in atoms trapped near a waveguide. Optica {\bf 3}, 725 (2016).
\bibitem{Ruostekoski2016} J. Ruostekoski and J. Javanainen, Emergence of correlated optics in one-dimensional waveguides for classical and quantum atomic gases. Phys. Rev. Lett. {\bf 117}, 143602 (2016).
\bibitem{Ruostekoski2017} J. Ruostekoski and J. Javanainen, Arrays of strongly coupled atoms in a one-dimensional waveguide. Phys. Rev. A {\bf 96}, 033857 (2017).
\bibitem{Kien2017} F. Le Kien and A. Rauschenbeutel, A. Nanofiber-mediated chiral radiative coupling between two atoms. Phys. Rev. A {\bf 95}, 023838 (2017).
\bibitem{Gardiner1993} C. W. Gardiner, Driving a quantum system with the output field from another driven quantum system, Phys. Rev. Lett. {\bf 70} 2269 (1993). 
\bibitem{Carmichael1993} H. J. Carmichael, Quantum trajectory theory for cascaded open systems, Phys. Rev. Lett. {\bf 70} 2273 (1993).
\bibitem{Stannigel2012} K. Stannigel, P. Rabl, and P. Zoller, Driven-dissipative preparation of entangled states in cascaded quantum-optical networks, New J. Phys. {\bf 14}, 063014 (2012).
\bibitem{Downing2020} C. A. Downing, J. C. López Carreño, A. I. Fernández-Domínguez, and E. del Valle, Asymmetric coupling between two quantum emitters, Phys. Rev. A {\bf 102}, 013723 (2020). 
\bibitem{Mitsch2014} R. Mitsch, C. Sayrin, B. Albrecht, P. Schneeweiss, and A. Rauschenbeutel, Quantum state-controlled directional spontaneous emission of photons into a nanophotonic waveguide, Nat. Commun. {\bf 5}, 5713 (2014).
\bibitem{Samutpraphoot2020} P. Samutpraphoot, T. Dordevi\ifmmode \acute{c}\else \'{c}\fi{}, P. L. Ocola, H. Bernien, C. Senko, V. Vuletić, and M. D. Lukin, Strong Coupling of Two Individually Controlled Atoms via a Nanophotonic Cavity, Phys. Rev. Lett. {\bf 124}, 063602 (2020). 
\bibitem{Liu2023} Y. Liu, Z. Wang, P. Yang, Q. Wang, Q. Fan, S. Guan, G. Li, P. Zhang, and T. Zhang, Realization of Strong Coupling between Deterministic Single-Atom Arrays and a High-Finesse Miniature Optical Cavity, Phys. Rev. Lett. {\bf 130}, 173601 (2023). 
\bibitem{Li2023} Z. Li, S. Colombo, C. Shu, G. Velez, S. Pilatowsky-Cameo, R. Schmied, S. Choi, M. Lukin, E. Pedrozo-Peñafiel, V. Vuletić, Improving metrology with quantum scrambling, Science {\bf 380}, 1381 (2023). 
\bibitem{Dordevic2021} T. Dordevi\ifmmode \acute{c}\else \'{c}\fi{}, P. Samutpraphoot, P. L. Ocola, H. Bernien, B. Grinkemeyer, I. Dimitrova, V. Vuleti\'{c}, and M. D. Lukin, Entanglement transport and a nanophotonic interface for atoms in optical tweezers, Science {\bf 373}, 1511 (2021). 
\bibitem{Cooper2024} E. S. Cooper, P. Kunkel, A. Periwal, and M. Schleier-Smith, Graph states of atomic ensembles engineered by photon-mediated entanglement, Nat. Phys. {\bf 20}, 770 (2024). 
\bibitem{Chien2024} C.-H. Chien, S. Goswami, C.-C. Wu, W.-S. Hiew, Y.-C. Chen, and H. H. Jen, Generating Graph States in an Atom-Nanophotonic Interface, Quantum Sci. Technol. {\bf 9}, 025020 (2024). 
\bibitem{Hiew2023} W. S. Hiew, and H. H. Jen, State Carving in a Chirally-Coupled Atom-Nanophotonic Cavity, New J. Phys. {\bf 25}, 093018 (2023). 
\bibitem{Henriet2019} L. Henriet, J. S. Douglas, D. E. Chang, and A. Albrecht, Critical open-system dynamics in a one-dimensional optical-lattice clock, Phys. Rev. A {\bf 99}, 023802 (2019).
\bibitem{Kornovan2019}  D. F. Kornovan, N. V. Corzo, J. Laurat, and A. S. Sheremet, Extremely subradiant states in a periodic one-dimensional atomic array, Phys. Rev. A {\bf 100}, 063832 (2019). 
\bibitem{Jen2020_subradiance} H. H. Jen, M.-S. Chang, G.-D. Lin, and Y.-C. Chen, Subradiance dynamics in a singly excited chirally coupled atomic chain, Phys. Rev. A {\bf 101}, 023830 (2020).
\bibitem{Ke2019} Y. Ke, A. V. Poshakinskiy, C. Lee, Y. S. Kivshar, and A. N. Poddubny, Inelastic Scattering of Photon Pairs in Qubit Arrays with Subradiant States, Phys. Rev. Lett. {\bf 123}, 253601 (2019).
\bibitem{Kumlin2020} J. Kumlin, K. Kleinbeck, N. Stiesdal, H. Busche, S. Hofferberth, and H. P. Büchler, Nonexponential decay of a collective excitation in an atomic ensemble coupled to a one-dimensional waveguide, Phys. Rev. A {\bf 102}, 063703 (2020). 
\bibitem{Berman2020} P. R. Berman, Theory of two atoms in a chiral waveguide, Phys. Rev. A {\bf 101}, 013830 (2020). 
\bibitem{Pivovarov2021} V. A. Pivovarov, L. V. Gerasimov, J. Berroir, T. Ray, J. Laurat, A. Urvoy, and D. V. Kupriyanov, Single collective excitation of an atomic array trapped along a waveguide: A study of cooperative emission for different atomic chain configurations, Phys. Rev. A {\bf 103}, 043716 (2021). 
\bibitem{Jen2021_bound} H. H. Jen, Bound and subradiant multiatom excitations in an atomic array with nonreciprocal couplings, Phys. Rev. A {\bf 103}, 063711 (2021). 
\bibitem{Zhang2020_bound} Y.-X. Zhang, C. Yu, and K. Mølmer, Subradiant bound dimer excited states of emitter chains coupled to a one dimensional waveguide, Phys. Rev. Research {\bf 2}, 013173 (2020).
\bibitem{Zhong2020} J. Zhong, N. A. Olekhno, Y. Ke, A. V. Poshakinskiy, C. Lee, Y. S. Kivshar, and A. N. Poddubny, Photon-Mediated Localization in Two-Level Qubit Arrays, Phys. Rev. Lett. {\bf 124}, 093604 (2020).
\bibitem{Mirza2017} I. M. Mirza, J.G. Hoskins, and J. C. Schotland, Chirality, band structure, and localization in waveguide quantum electrodynamics, Phys. Rev. A {\bf 96}, 053804 (2017). 
\bibitem{Jen2020_disorder} H. H. Jen, Disorder-assisted excitation localization in chirally coupled quantum emitters, Phys. Rev. A {\bf 102}, 043525 (2020). 
\bibitem{Jen2021_crossover} H. H. Jen and J-S You, Crossover from a delocalized to localized atomic excitation in an atom–waveguide interface, J. Phys. B: At. Mol. Opt. Phys. {\bf 54}, 105002 (2021). 
\bibitem{Jen2022_correlation} H. H. Jen, Quantum correlations of localized atomic excitations in a disordered atomic chain, Phys. Rev. A {\bf 105}, 023717 (2022). 
\bibitem{Schrinski2022} B. Schrinski and A. S. Sørensen, Polariton dynamics in one-dimensional arrays of atoms coupled to waveguides, New J. Phys. {\bf 24}, 123023 (2022). 
\bibitem{Tonks1936} L. Tonks, The complete equation of state of one, two and three-dimensional gases of hard elastic spheres, Phys. Rev. {\bf 50}, 955 (1936). 
\bibitem{Girardeau1960} M. Girardeau, Relationship between systems of impenetrable bosons and fermions in one dimension, J. Math. Phys. {\bf 1}, 516 (1960). 
\bibitem{Iversen2022} O. A. Iversen and T. Pohl, Self-ordering of individual photons in waveguide QED and Rydberg-atom arrays, Phys. Rev. Research {\bf 4}, 023002 (2022).
\bibitem{Liao2020} Z. Liao , Y. Lu, and M. S. Zubairy, Multiphoton pulses interacting with multiple emitters in a one-dimensional waveguide, Phys. Rev. A {\bf 102}, 053702 (2020). 
\bibitem{Pennetta2022} R. Pennetta, M. Blaha, A. Johnson, D. Lechner, P. Schneeweiss, J. Volz, and A. Rauschenbeutel, Collective Radiative Dynamics of an Ensemble of Cold Atoms Coupled to an Optical Waveguide, Phys. Rev. Lett. {\bf 128}, 073601 (2022).
\bibitem{Pennetta2022_2} R. Pennetta, D. Lechner, M. Blaha , A. Rauschenbeutel, P. Schneeweiss, and J. Volz, Observation of Coherent Coupling between Super- and Subradiant States of an Ensemble of Cold Atoms Collectively Coupled to a Single Propagating Optical Mode, Phys. Rev. Lett. {\bf 128}, 203601 (2022).
\bibitem{Zhang2020} Y.-X. Zhang and K. Mølmer, Subradiant Emission from Regular Atomic Arrays: Universal Scaling of Decay Rates from the Generalized Bloch Theorem, Phys. Rev. Lett. {\bf 125}, 253601 (2020).
\bibitem{Zhang2022} Y.-X. Zhang and K. Mølmer, Free-Fermion Multiply Excited Eigenstates and Their Experimental Signatures in 1D Arrays of Two-Level Atoms, Phys. Rev. Lett. {\bf 128}, 093602 (2022).
\bibitem{Chen2023} C.-C. Chen, Y.-C. Wang, C.-C. Wang, and H. H. Jen, Chiral-coupling-assisted refrigeration in trapped ions, J. Phys. B: At. Mol. Opt. Phys. {\bf 56}, 105502 (2023).  
\bibitem{Wang2022} C.-C. Wang, Y.-C. Wang, C.-H. Wang, C.-C. Chen, and H. H. Jen, Superior dark-state cooling via nonreciprocal couplings in trapped atoms”, New J. of Phys. {\bf 24}, 113020 (2022). 
\bibitem{Pino2021} J. M. Pino, J. M. Dreiling, C. Figgatt, J. P. Gaebler, S. A. Moses, M. S. Allman, C. H. Baldwin, M. Foss-Feig, D. Hayes, K. Mayer, {\it et al.}, Demonstration of the trapped-ion quantum CCD computer architecture, Nature {\bf 592}, 209 (2021). 
\bibitem{Barik2018} S. Barik, A. Karasahin, C. Flower, T. Cai, H. Miyake, W.DeGottardi, M. Hafezi, and E. Waks, A Topological Quantum Optics Interface, Science {\bf 359}, 666 (2018).
\bibitem{Bello2019} M. Bello, G. Platero, J. I. Cirac, and A. González-Tudela, Unconventional Quantum Optics in Topological Waveguide QED, Sci. Adv. {\bf 5}, eaaw0297 (2019).
\bibitem{Krantz2019} P. Krantz, M. Kjaergaard, F. Yan, T. P. Orlando, S.Gustavsson, and W. D. Oliver, A Quantum Engineer’sGuide to Superconducting Qubits, Appl. Phys. Rev. {\bf 6}, 021318 (2019).
\bibitem{Shi2018} T. Shi, Y.-H. Wu, A. González-Tudela, and J. I. Cirac,Effective Many-Body Hamiltonians of Qubit-Photon BoundStates, New J. Phys. {\bf 20}, 105005 (2018).
\bibitem{Sundaresan2019} N. M. Sundaresan, R. Lundgren, G. Zhu, A. V. Gorshkov,and A. A. Houck, Interacting Qubit-Photon Bound Stateswith Superconducting Circuits, Phys. Rev. X {\bf 9}, 011021(2019).
\bibitem{Su1979} W. P. Su, J. R. Schrieffer, and A. J. Heeger, Solitons inPolyacetylene, Phys. Rev. Lett. {\bf 42}, 1698 (1979).
\bibitem{Diehl2008} S. Diehl, A. Micheli, A. Kantian, B. Kraus, H. P. B\"{u}chler, and P. Zoller, Quantum states and phases in driven open quantum systems with cold atoms, Nat. Phys. {\bf 4}, 878 (2008).
\bibitem{Kraus2008} B. Kraus, H. P. B\"{u}chler, S. Diehl, A. Kantian, A. Micheli, and P. Zoller, Preparation of entangled states by quantum Markov processes, Phys. Rev. A {\bf 78}, 042307 (2008).
\bibitem{Verstraete2009} F. Verstraete, M. M. Wolf, and J. I. Cirac, Quantum computation and quantum-state engineering driven by dissipation, Nat. Phys. {\bf 5}, 633 (2009).
\bibitem{Diehl2010} S. Diehl, A. Tomadin, A. Micheli, R. Fazio, and P. Zoller, Dynamical Phase Transitions and Instabilities in Open Atomic Many-Body Systems, Phys. Rev. Lett. {\bf 105}, 015702 (2010).
\bibitem{Diehl2011} S. Diehl, E. Rico, M. A. Baranov, and P. Zoller, Topology by dissipation in atomic quantum wires, Nat. Phys. {\bf 7}, 971 (2011).
\bibitem{Buonaiuto2019} G. Buonaiuto, R. Jones, B. Olmos and I. Lesanovsky, Dynamical creation and detection of entangled many-body states in a chiral atom chain, New J. Phys. {\bf 21}, 113021 (2019). 
\bibitem{Kornovan2016} D. F. Kornovan, A. S. Sheremet, and M. I. Petrov, Collective polaritonic modes in an array of two-level quantum emitters coupled to an optical nanofiber, Phys. Rev. B {\bf 94}, 245416 (2016).
\bibitem{Birkl1995} G. Birkl, M. Gatzke, I. H. Deutsch, S. L. Rolston, and W. D. Phillips, Bragg Scattering from Atoms in Optical Lattices, Phys. Rev. Lett. {\bf 75}, 2823 (1995).
\bibitem{Schilke2011} A. Schilke, C. Zimmermann, P. W. Courteille, and W. Guerin, Photonic Band Gaps in One-Dimensionally Ordered Cold Atomic Vapors, Phys. Rev. Lett. {\bf 106}, 223903 (2011).
\bibitem{Olmos2021} B. Olmos, C. Liedl, I. Lesanovsky, and P. Schneeweiss, Bragg condition for scattering into a guided optical mode, Phys. Rev. A {\bf 104}, 043517 (2021).
\bibitem{Jen2020_steady} H. H. Jen, Steady-state phase diagram of a weakly driven chiral-coupled atomic chain, Phys. Rev. Research {\bf 2}, 013097 (2020).
\bibitem{Voorden2020} B. van Voorden, J. Minář, and K. Schoutens, Quantum many-body scars in transverse field Ising ladders and beyond, Phys. Rev. B {\bf 101}, 220305(R) (2020).
\bibitem{Voorden2021} B. van Voorden, M. Marcuzzi, K. Schoutens, and J. Minář, Disorder enhanced quantum many-body scars in Hilbert hypercubes, Phys. Rev. B {\bf 103}, L220301 (2021).
\bibitem{Turner2018_1} C. J. Turner, A. A. Michailidis, D. A. Abanin, M. Serbyn, and Z. Papić, Weak ergodicity breaking from quantum many-body scars, Nat. Phys. {\bf 14}, 745 (2018).
\bibitem{Turner2018_2} C. J. Turner, A. A. Michailidis, D. A. Abanin, M. Serbyn, and Z. Papić, Quantum scarred eigenstates in a Rydberg atom chain: Entanglement, breakdown of thermalization, and stability to perturbations, Phys. Rev. B {\bf 98}, 155134 (2018).
\bibitem{Lin2019} C.-J. Lin and O. I. Motrunich, Exact Quantum Many-Body Scar States in the Rydberg-Blockaded Atom Chain, Phys. Rev. Lett. {\bf 122}, 173401 (2019).
\bibitem{Heller1984} E. J. Heller, Bound-State Eigenfunctions of Classically Chaotic Hamiltonian Systems: Scars of Periodic Orbits, Phys. Rev. Lett. {\bf 53}, 1515 (1984).
\bibitem{Kinoshita2006} T. Kinoshita, T. Wenger, and D. S. Weiss, A quantum Newton's cradle, Nature {\bf 440}, 900 (2006).
\bibitem{Tsaur1998} G.-y. Tsaur and J. Wang, Population Diffusion and Equipartition in Quantum Systems of Many Degrees of Freedom, Phys. Rev. Lett. {\bf 80}, 3682 (1998). 
\bibitem{Srednicki1999} M. Srednicki, The approach to thermal equilibrium in quantized chaotic systems, J. Phys. A: Math. Gen. {\bf 32}, 1163 (1999). 
\bibitem{Bernien2017} H. Bernien, S. Schwartz, A. Keesling, H. Levine, A. Omran, H. Pichler, S. Choi, A. S. zibrov, M. Endres, M. Grenier, V. Vuletić, and M.D. Lukin, Probing many-body dynamics on a $51$-atom quantum simulator, Nature {\bf 551}, 579 (2017).
\bibitem{Choi2017} S. Choi, J. Choi, R. Landig, G. Kucsko, H. Zhou, J. Isoya, F. Jelezko, S. Onoda, H. Sumiya, V. Khemani, C. von Keyserlingk, N. Y. Yao, E. Demler, and M. D. Lukin, Observation of discrete time-crystalline order in a disordered dipolar many-body system, Nature {\bf 543}, 221 (2017). 
\bibitem{Chung2024} S.-H. Chung, I G. N. Y. Handayana, Y.-L. Tsao, C.-C. Wu, G.-D. Lin, H. H. Jen, Steady-state phases and interaction-induced depletion in a driven-dissipative chirally-coupled dissimilar atomic array, Phys. Rev. Research {\bf 6}, 023232 (2024). 
\bibitem{Handayana2024} I G. N. Y. Handayana, C.-C. Wu, S. Goswami, Y.-C. Chen, H. H. Jen, Atomic excitation trapping in dissimilar chirally-coupled atomic arrays, Phys. Rev. Research {\bf 6}, 013320 (2024). 
\bibitem{Anderson1958} P. W. Anderson, Absence of diffusion in certain random lattices, Phys. Rev. {\bf 109}, 1492 (1958).
\bibitem{Clement2005} D. Clèment, A. F. Varón, M. Hugbart, J. A. Retter, P. Bouyer, L. Sanchez-Palencia, D. M. Gangardt, G. V. Shlyapnikov, and A. Aspect, Suppression of transport of an interacting elongated bose-einstein condensate in a random potential, Phys. Rev. Lett. {\bf 95}, 170409 (2005).
\bibitem{Evers2008} F. Evers and A. D. Mirlin, Anderson transitions, Rev. Mod. Phys. {\bf 80}, 1355 (2008). 
\bibitem{Wu2024}  C.-C. Wu, K.-T. Lin, I. G. N. Y. Handayana, C.-H. Chien, S. Goswami, G.-D. Lin, Y.-C. Chen, and H. H. Jen, Atomic excitation delocalization at the clean to disordered interface in a chirally-coupled atomic array, Phys. Rev. Research {\bf 6}, 013159 (2024). 
\bibitem{Luitz2017} D. J. Luitz, F. Huveneers, and W. De Roeck, How a small quantum bath can thermalize long localized chains, Phys. Rev. Lett. {\bf 119}, 150602 (2017).
\bibitem{Roeck2017} W. De Roeck and F.Huveneers, Stability and instability towards delocalization in many-body localization systems, Phys. Rev. B {\bf 95}, 155129 (2017). 
\bibitem{Thiery2018} T. Thiery, F. Huveneers, M. Müller, and W. De Roeck, Many-body delocalization as a quantum avalanche, Phys. Rev. Lett. {\bf 121}, 140601 (2018). 
\bibitem{Morningstar2022} A. Morningstar, L. Colmenarez, V. Khemani, D. J. Luitz, and D. A. Huse, Avalanches and many-body resonances in many-body localized systems, Phys. Rev. B {\bf 105}, 174205 (2022). 
\bibitem{Sels2022} D. Sels, Bath-induced delocalization in interacting disordered spin chains, Phys. Rev. B {\bf 106}, L020202 (2022). 
\bibitem{Leonard2023} J. Léonard, S. Kim, M. Rispoli, A. Lukin, R. Schittko, J. Kwan, E. Demler, D. Sels, and M. Greiner, Probing the onset of quantum avalanches in a many-body localized system, Nat. Phys. {\bf 19}, 481 (2023).
\bibitem{Luschen2017}  H. P. Lüschen, P. Bordia, S. S. Hodgman, M. Schreiber, S. Sarkar, A. J. Daley, M. H. Fischer, E. Altman, I. Bloch, and U. Schneider, Signatures of Many-Body Localization in a Controlled Open Quantum System, Phys. Rev. X {\bf 7}, 011034 (2017). 
\bibitem{Patti2021} T. L. Patti, D. S. Wild, E. Shahmoon, M. D. Lukin, and S. F. Yelin, Controlling Interactions between Quantum Emitters Using Atom Arrays, Phys. Rev. Lett. {\bf 126}, 223602 (2021).
\bibitem{Buckley-Bonanno2022} S. Buckley-Bonanno, S. Ostermann, O. Rubies-Bigorda, T. L. Patti, and S. F. Yelin, Optimized geometries for cooperative photon storage in an impurity coupled to a two-dimensional atomic array, Phys. Rev. A {\bf 106}, 053706 (2022). 
\bibitem{Mann2022} C.-R. Mann and E. Mariani, Topological transitions in arrays of dipoles coupled to a cavity waveguide, Phys. Rev. Research {\bf 4}, 013078 (2022).
\bibitem{Wang2018} B. X. Wang and C. Y. Zhao, Topological photonic states in one-dimensional dimerized ultracold atomic chains, Phys. Rev. A {\bf 98}, 023808 (2018).
\bibitem{Poshakinskiy2021} A. V. Poshakinskiy, J. Zhong, Y. Ke, N. A. Olekhno, C. Lee, Y. S. Kivshar, and A. N. Poddubny, Quantum Hall phases emerging from atom–photon interactions, npj Quantum Information {\bf 7}, 34 (2021). 
\bibitem{Ashida2020} Y. Ashida, Z. Gong, and M. Ueda, Non-Hermitian physics, Adv. Phys. {\bf 69}, 3 (2020).
\bibitem{El-Ganainy2018} R. El-Ganainy, K. G. Makris, M. Khajavikhan, Z. H. Musslimani, S. Rotter, and D. N. Christodoulides, Non-Hermitian physics and PT symmetry, Nat. Phys. {\bf 14}, 11 (2018).
\bibitem{Bergholtz2021} E. J. Bergholtz, J. C. Budich, and F. K. Kunst, Exceptional topology of non-Hermitian systems, Rev. Mod. Phys. {\bf 93}, 015005 (2021).
\bibitem{Lee2016} T. E. Lee, Anomalous Edge State in a Non-Hermitian Lattice, Phys. Rev. Lett. {\bf 116}, 133903 (2016).
\bibitem{Yao2018} S. Yao and Z. Wang, Edge states and topological invariants of non-Hermitian systems, Phys. Rev. Lett. \textbf{121}, 086803 (2018).
\bibitem{MartinezAlvarez2018}V. M. Martinez Alvarez, J. E. Barrios Vargas, and L.E. F. Foa Torres, Non-Hermitian robust edge states in one dimension: anomalous localization and eigenspace condensation at exceptional points, Phys. Rev. B \textbf{97}, 121401(R) (2018).
\bibitem{Kunst2018} F. K. Kunst, E. Edvardsson, J. C. Budich, and E. J. Bergholtz, Biorthogonal bulk boundary correspondence in non-hermitian systems, Phys. Rev. Lett. \textbf{121}, 026808 (2018).
\bibitem{Lee2019} C. H. Lee and R. Thomale, Anatomy of skin modes and topology in non-Hermitian systems, Phys. Rev. B {\bf 99}, 201103(R) (2019). 
\bibitem{Borgnia2020} D. S. Borgnia, A. J. Kruchkov, and R.-J. Slager, Non-hermitian boundary modes and topology, Phys. Rev. Lett. \textbf{124}, 056802 (2020).
\bibitem{Okuma2020} N. Okuma, K. Kawabata, K. Shiozaki, and M. Sato, Topological origin of non-hermitian skin effects, Phys. Rev. Lett. \textbf{124}, 086801 (2020).
\bibitem{Zhang2020_skin} K. Zhang, Z. Yang, and C. Fang, Correspondence between winding numbers and skin modes in non-hermitian systems, Phys. Rev. Lett. \textbf{125}, 126402 (2020).
\bibitem{Wang2024_skin} H. Wang, J. Zhong, and S. Fan, Non-Hermitian photonic band winding and skin effects: a tutorial, Advances in Optics and Photonics {\bf 16}, 659 (2024). 
\bibitem{Zhang2022_skin} K. Zhang, Z. Yang, and C. Fang, Universal non-Hermitian skin effect in two and higher dimensions. Nat. Commun. {\bf 13}, 2496 (2022).
\bibitem{Wang2022_skin} Y.-C. Wang, J.-S. You, and H. H. Jen, A non-Hermitian optical atomic mirror, Nat. Commun. {\bf 13}, 4598 (2022). 
\bibitem{Hatano1996} N. Hatano and D. R. Nelson, Localization Transitions in Non-Hermitian Quantum Mechanics, Phys. Rev. Lett. {\bf 77}, 570 (1996).
\bibitem{Wang2023_scaling} Y.-C. Wang, H. H. Jen, and J.-S. You, Scaling laws for non-Hermitian skin effect with long-range couplings, Phys. Rev. B {\bf 108}, 085418 (2023).  
\bibitem{Wang2023_skin} Y.-C. Wang, K. Suthar, H. H. Jen, Y.-T. Hsu, J.-S. You, Non-Hermitian skin effects on thermal and many-body localized phases, Phys. Rev. B {\bf 107}, L220205 (2023). 
\bibitem{Abanin2019} D. A. Abanin, E. Altman, I. Bloch, and M. Serbyn, Colloquium: Many-body localization, thermalization, and entanglement, Rev. Mod. Phys. {\bf 91}, 021001 (2019).
\bibitem{Hamazaki2019} R. Hamazaki, K. Kawabata, and M. Ueda, Non-Hermitian Many-Body Localization, Phys. Rev. Lett. {\bf 123}, 090603 (2019).
\bibitem{Zhen2015} B. Zhen,  C. W. Hsu, Y. Igarashi, L. Lu, I. Kaminer, A. Pick, S.-L. Chua, J. D. Joannopoulos, and M. Soljačić, Spawning rings of exceptional points out of Dirac cones, Nature \textbf{525}, 354 (2015).
\bibitem{Zhou2018} H. Zhou,  C. Peng, Y. Yoon, C. W. Hsu, K. A. Nelson, L. Fu, J. D. Joannopoulos, M. Soljačić, and B. Zhen, Observation of bulk Fermi arc and polarization half charge from paired exceptional points, Science \textbf{359}, 1009 (2018).
\bibitem{Lee2018} C. H. Lee, S. Imhof, C. Berger, F. Bayer, J. Brehm, L. W. Molenkamp, T. Kiessling, and R. Thomale, Topolectrical circuits, Commun. Phys. {\bf 1}, 39 (2018).
\bibitem{Hofmann2020} T. Hofmann, T. Helbig, F. Schindler, N. Salgo, M. Brzezińska, M. Greiter, T. Kiessling, D. Wolf, A. Vollhardt, A. Kabaši, C. H. Lee, A. Bilušić, R. Thomale, and T. Neupert, Reciprocal skin effect and its realization in a topolectrical circuit, Phys. Rev. Res. \textbf{2}, 023265 (2020).
\bibitem{Wang2024_nexus} C. Wang, N. Li, J. Xie, C. Ding, Z. Ji, L. Xiao, S. Jia, B. Yan, Y. Hu, and Y. Zhao, Exceptional Nexus in Bose-Einstein Condensates with Collective Dissipation, Phys. Rev. Lett. {\bf 132}, 253401 (2024).
\bibitem{Holzinger2024_2} R. Holzinger, O. Rubies-Bigorda, S. F. Yelin, and H. Ritsch, Symmetry based efficient simulation of dissipative quantum many-body dynamics in subwavelength quantum emitter arrays, arXiv: 2409.02790. 
\bibitem{Marques2021} Y. Marques, I. A. Shelykh, and I. V. Iorsh, Bound Photonic Pairs in 2D Waveguide Quantum Electrodynamics, Phys. Rev. Lett. {\bf 127}, 273602 (2021). 
\bibitem{Shahmoon2020} E. Shahmoon, M. D. Lukin, and S. F. Yelin, Quantum optomechanics of a two-dimensional atomic array, Phys. Rev. A {\bf 101}, 063833 (2020). 
\bibitem{Aspelmeyer2014} M. Aspelmeyer, T. J. Kippenberg, and F. Marquardt, Cavity optomechanics, Rev. Mod. Phys. {\bf 86}, 1391 (2014). 
\bibitem{Brantut2013} J.-P. Brantut, C. Grenier, J. Meineke, D. Stadler, S. Krinner, C. Kollath, T. Esslinger, and A. Georges, A thermoelectric heat engine with ultracold atoms, Science {\bf 342}, 713 (2013). 
\bibitem{Carollo2020} F. Carollo, F. M. Gambetta, K. Brandner, J. P. Garrahan, and I. Lesanovsky, Nonequilibrium quantum many-body Rydberg atom engine, Phys. Rev. Lett. {\bf 124}, 170602 (2020).
\bibitem{Myers2022} N. M. Myers, F. J. Peña, O. Negrete, P. Vargas, G. De Chiara, and S. Deffner, Boosting engine performance with Bose–Einstein condensation, New J. Phys. {\bf 24}, 025001 (2022). 
\bibitem{Koch2023} J. Koch, K. Menon, E. Cuestas, S. Barbosa, E. Lutz, T. Fogarty, T. Busch, and A. Widera, A quantum engine in the BEC–BCS crossover, Nature {\bf 621}, 723 (2023). 
\bibitem{Nautiyal2024} V. V. Nautiyal, R. S. Watson, and K. V. Kheruntsyan, A finite-time quantum Otto engine with tunnel coupled one-dimensional Bose gases, New J. Phys. {\bf 26}, 063033 (2024). 
\bibitem{Estrada2024} J. A. Estrada, F. Mayo, A. J. Roncaglia, and P. D. Mininni, Quantum engines with interacting Bose-Einstein condensates, Phys. Rev. A {\bf 109}, 012202 (2024). 
\bibitem{Feyisa2024} C. G. Feyisa and H. H. Jen, Trapped-atom Otto engine with light-induced dipole–dipole interactions, New J. Phys. {\bf 26}, 093039 (2024). 
\bibitem{Kien2022} F. Le Kien, L. Ruks, S. Nic Chormaic, and T. Busch, Coupling between guided modes of two parallel nanofibers, New J. Phys. {\bf 22}, 123007 (2020).
\bibitem{Kien2024} K. Jain, L. Ruks, F. le Kien, and T. Busch, Strong dipole-dipole interactions via enhanced light-matter coupling in composite nanofiber waveguides, Phys. Rev. Research {\bf 6},033311(2024). 
\bibitem{Sorensen2003} A. S. Sørensen and K. Mølmer, Probabilistic Generation of Entanglement in Optical Cavities, Phys. Rev. Lett. {\bf 90}, 127903 (2003). 
\bibitem{Chen2015} W. Chen, J. Hu, Y. Duan, B. Braverman, H. Zhang, and V. Vuletić, Carving Complex Many-Atom Entangled States by Single-Photon Detection, Phys. Rev. Lett. {\bf 115}, 250502 (2015).
\bibitem{Welte2017} S. Welte, B. Hacker, S. Daiss, S. Ritter, and G. Rempe, Cavity Carving of Atomic Bell States, Phys. Rev. Lett. {\bf 118}, 210503 (2017).
\bibitem{Briegel2001} H. J. Briegel and R. Raussendorf, Persistent Entanglement in Arrays of Interacting Particles, Phys. Rev. Lett. {\bf 86}, 910 (2001). 
\bibitem{Hein2004} M. Hein, J. Eisert, and H. J. Briegel, Multi-party entanglement in graph states, Phys. Rev. A {\bf 69}, 062311 (2004).
\bibitem{Lanyon2013} B. P. Lanyon, P. Jurcevic, M. Zwerger, C. Hempel, E. A. Martinez, W. D\"{u}r, H. J. Briegel, R. Blatt, and C. F. Roos, Phys. Rev. Lett. {\bf 111}, 210501 (2013).
\bibitem{Gong2019} M. Gong, M.-C. Chen, Y. Zheng, S. Wang, C. Zha, H. Deng, Z. Yan, H. Rong, Y. Wu, S. Li, {\it et al.}, Phys. Rev. Lett. {\bf 122}, 110501 (2019).
\bibitem{Cao2023} S. Cao, B. Wu, F. Chen, M. Gong, Y. Wu, Y. Ye, C. Zha, H. Qian, C. Ying, S. Guo, {\it et al.}, Nature {\bf 619}, 738 (2023).
\bibitem{Raussendorf2001} R. Raussendorf and H. J. Briegel, A one-way quantum computer. Phys. Rev. Lett. {\bf 86}, 5188 (2001).
\bibitem{Raussendorf2003} R. Raussendorf, D. E. Browne, and H. J. Briegel, Measurement-based quantum computation on cluster states, Phys. Rev. A {\bf 68}, 022312 (2003).
\bibitem{Walther2005} P. Walther, K. J. Resch, T. Rudolph, E. Schenck, H. Weinfurter, V. Vedral, M. Aspelmeyer, and A. Zeilinger, Experimental one-way quantum computing, Nature {\bf 434}, 169 (2005). 
\bibitem{Briegel2009} H. J. Briegel, D. E. Browne, W. Dür, R. Raussendorf, and M. Van den Nest, Measurement-based quantum computation, Nat. Phys. {\bf 5}, 19 (2009).
\bibitem{Bluvstein2022} D. Bluvstein, H. Levine, G. Semeghini, T. T. Wang, S. Ebadi, M. Kalinowski, A. Keesling, N. Maskara, H. Pichler, M. Greiner, V. Vuletić, and M. D. Lukin, A quantum processor based on coherent transport of entangled atom arrays, Nature {\bf 604}, 451 (2022).
\bibitem{Bluvstein2024} D. Bluvstein, S. J. Evered, A. A. Geim, S. H. Li, H. Zhou, T. Manovitz, S. Ebadi, M. Cain, M. Kalinowski, D. Hangleiter, J. P. B. Ataides, N. Maskara, I. Cong, X. Gao, P. S. Rodriguez, T. Karolyshyn, G. Semeghini, M. J. Gullans, M. Greiner, V. Vuletić, and M. D. Lukin, Logical quantum processor based on reconfigurable atom arrays, Nature {\bf 626}, 58 (2024).
\bibitem{Kiesel2005} N. Kiesel, C. Schmid, U. Weber, G. T\'{o}th, O. G\"{u}hne, R. Ursin, and H. Weinfurter, Experimental Analysis of a Four-Qubit Photon Cluster State, Phys. Rev. Lett. {95}, 210502 (2005). 
\bibitem{Lu2007} C.-Y. Lu, X.-Q. Zhou, O. Gühne, W.-B. Gao, J. Zhang, Z.-S. Yuan, A. Goebel, T. Yang, J.-W. Pan, Experimental entanglement of six photons in graph states, Nat. Phys. {\bf 3}, 91 (2007).
\bibitem{Tokunaga2008} Y. Tokunaga, S. Kuwashiro, T. Yamamoto, M. Koashi, N. Imoto, Generation of high-fidelity four-photon cluster state and quantum-domain demonstration of one-way quantum computing, Phys. Rev. Lett. {\bf 100}, 210501 (2008).
\bibitem{Schwartz2016} I. Schwartz, D. Cogan, E. R. Schmidgall, Y. Don, L. Gantz, O. Kenneth, N. H. Lindner, and D. Gershoni, Deterministic generation of a cluster state of entangled photons, Science {\bf 354}, 434 (2016). 
\bibitem{Larsen2019} M. V. Larsen, X. Guo, C. R. Breum, J. S. Neergaard-Nielsen, and U. L. Andersen, Deterministic generation of a two-dimensional cluster state, Science {\bf 366}, 369 (2019). 
\bibitem{Asavanant2019} W. Asavanant, Y. Shiozawa, S. Yokoyama, B. Charoensombutamon, H. Emura, R. N. Alexander, S. Takeda, J.-i. Yoshikawa, N. C. Menicucci, H. Yonezawa, and A. Furusawa, Generation of time-domain-multiplexed two-dimensional cluster state, Science {\bf 366}, 373 (2019). 
\bibitem{Larsen2021} M. V. Larsen, X. Guo, C. R. Breum, J. S. Neergaard-Nielsen, and U. L. Andersen, Deterministic multi-mode gates on a scalable photonic quantum computing platform, Nat. Phys. {\bf 17}, 1018 (2021). 
\bibitem{Thomas2022} P. Thomas, L. Ruscio, O. Morin, and G. Rempe, Efficient generation of entangled multiphoton graph states from a single atom, Nature {\bf 608}, 677 (2022). 
\bibitem{Yang2022} C.-W. Yang, Y. Yu, J. Li, B. Jing, X.-H. Bao, and J.-W. Pan, Sequential generation of multiphoton entanglement with a Rydberg superatom, Nat. Photon. {\bf 16}, 658 (2022). 
\bibitem{Kaufman2012} A. M. Kaufman, B. J. Lester, and C. A. Regal, Cooling a Single Atom in an Optical Tweezer to Its Quantum Ground State, Phys. Rev. X {\bf 2}, 041014 (2012).
\bibitem{Brown2019} M. O. Brown, T. Thiele, C. Kiehl, T.-W. Hsu, and C. A. Regal, Gray-Molasses Optical-Tweezer Loading: Controlling Collisions for Scaling Atom-Array Assembly, Phys. Rev. X {\bf 9}, 011057 (2019).  
\end{thebibliography}
\end{document}